\documentclass[reprint, 
amsmath,
amssymb, 
floatfix, 
aps,
prl,
superscriptaddress]{revtex4-2} 
\usepackage{graphicx}

\usepackage[utf8]{inputenc}
\usepackage{xcolor}
\definecolor{codegreen}{rgb}{0,0.6,0}
\definecolor{codegray}{rgb}{0.5,0.5,0.5}
\definecolor{codepurple}{rgb}{0.58,0,0.82}
\definecolor{backcolour}{rgb}{0.95,0.95,0.92}
\definecolor{urlblue}{HTML}{007bff}

\usepackage{bm}
\usepackage{bbm}
\usepackage{orcidlink}
\usepackage{microtype}
\usepackage{enumitem}
\usepackage{braket}
\usepackage{mathtools}

\usepackage{wasysym}
\usepackage{times}
\usepackage{appendix}
\usepackage[capitalise]{cleveref}
\input{macros.sty}
\usepackage{orcidlink} 
\usepackage[normalem]{ulem}

\hypersetup{
    colorlinks = true,
    linkcolor  = urlblue,
    citecolor  = urlblue,
    urlcolor   = urlblue,
}

\usepackage[mathlines]{lineno}

\usepackage{tabularx, booktabs} 

\usepackage{units} 
\usepackage{dsfont} 

\usepackage{changes}
\definechangesauthor[name=Valentin, color=blue]{VL} 

\begin{document}

\title{Topologically Protected Surface Altermagnetism on Antiferromagnets
}

\author{Valentin Leeb\orcidlink{0000-0002-7099-0682}}
\thanks{These authors contributed equally.}
\affiliation{Department of Physics, University of Zürich, Winterthurerstrasse 190, 8057 Zürich, Switzerland}
\author{Peru d'Ornellas\orcidlink{0000-0002-2349-0044}}
\thanks{These authors contributed equally.}
\affiliation{Donostia International Physics Center, P. Manuel de Lardizabal 4, 20018 Donostia-San Sebastian, Spain}
\affiliation{\small Universit\'e Grenoble Alpes, CNRS, Grenoble INP, Institut N\'eel, 38000 Grenoble, France}
\author{Fernando de Juan\orcidlink{0000-0001-6852-1484}}
\affiliation{Donostia International Physics Center, P. Manuel de Lardizabal 4, 20018 Donostia-San Sebastian, Spain}
\affiliation{IKERBASQUE, Basque Foundation for Science, Maria Diaz de Haro 3, 48013 Bilbao, Spain}
\author{Adolfo G. Grushin\orcidlink{0000-0001-7678-7100}}
\affiliation{Donostia International Physics Center, P. Manuel de Lardizabal 4, 20018 Donostia-San Sebastian, Spain}
\affiliation{IKERBASQUE, Basque Foundation for Science, Maria Diaz de Haro 3, 48013 Bilbao, Spain}
\affiliation{\small Universit\'e Grenoble Alpes, CNRS, Grenoble INP, Institut N\'eel, 38000 Grenoble, France}

\date{\today} 
	
\begin{abstract}
Altermagnetism (AM) and its associated spin-transport phenomena are typically linked to spin-split electronic band structures in bulk materials. However, the crystal surface has a reduced symmetry with respect to the bulk, which can induce AM at the surface of conventional antiferromagnets (AFMs)---a local effect which cannot be detected using bulk properties. In this work we define the symmetry conditions necessary for surface AM and show how it can be topologically protected, rendering it a robust effect. We provide a minimal model for one trivial and two topological examples of surface AM.
We show that the spin spectral density, accessible by spin- and angle-resolved photoemission spectroscopy, can exhibit a $d$-wave-like altermagnetic character at the surface, even when the full band structure is completely spin degenerate.
Our topological model describes the Dirac semimetal CuMnAs, which provides an existing realization of our theory.
Our results identify crystal surfaces as a platform to realize robust, topology- and symmetry-driven unconventional magnetism beyond the bulk classification of magnetic materials.
\end{abstract}

\maketitle

{\it Introduction---}%
Antiferromagnets (AFMs) are ordered, compensated magnets with a spin-degenerate electronic band structure. They are realized in a broad variety of materials ranging from elemental metals and transition-metal oxides to intermetallic compounds, layered van der Waals systems, and molecular magnets \cite{baltz_antiferromagnetic_2018}. Their ordering temperatures extend from the Kelvin scale to well above room temperature, underscoring the ubiquity and robustness of antiferromagnetic order. Altermagnets (AMs) have emerged as a separate symmetry-classified phase, defined by a non-relativistic combined spatial and time reversal symmetry \cite{hayami_bottom_2020,hayami_spontaneous_2020,hayami_momentum_2019,smejkal_2022_unconventional,smejkal_emerging_2022}. Both AM and AFMs are compensated magnets, however AMs realise a spin-split band structure which lies at the heart of various potential spintronic applications \cite{smejkal_beyond_2022,smejkal_emerging_2022}. In spite of the abundance of candidates, so far the search for AMs has revealed a surprising paucity of viable compounds beyond ab-initio predictions \cite{osumi_observation_2024,reimers_direct_2024}.

Here, we take a different path towards realizing AM: on the surface of conventional AFMs. 
From a symmetry perspective, the surface of a crystal is interesting because the presence of a boundary itself breaks any incompatible bulk space-group symmetries.
This symmetry reduction plays a central role in the emergence of surface states with properties forbidden in the bulk, such as in magnetic and topological systems \cite{davison_basic_1996, manchon_new_2015,jana_layering_2025}. 
Furthermore, in many materials, surface states benefit from topological protection, which ensures that their existence and gaplessness are robust against surface perturbations. 
In the context of AM, the majority of signatures---such as spin-transport and spectral properties at the Fermi level---are only measurable when the system is metallic. Thus, harnessing topology to produce surface states that are both metallic and altermagnetic on a conventional AFM is particularly enticing. Remonstrating this effect is the focus of this investigation.

As a concrete example for surface magnetism, consider the surface of a conventional AFM. Cutting the crystal along a plane such that only a single spin species exists at the surface, one generates a net surface magnetization---a surface ferromagnet with a spin-split surface state \cite{belashchenko_equilibrium_2010,chikina_strong_2014,nichols_emerging_2016}. Similarly, an AFM with non-vanishing surface octupolar moment exhibits an \emph{alter}magnetization \cite{jaeschke-ubiergo_atomic_2025,dornellas_altermagnetism_2025}. 
Although surface effects have been proposed as a mechanism for engineering altermagnetism in 3D AFMs \cite{mazin_induced_2023,zhou_transition_2025, meier_antialtermagnetism_2025}, research has focused so far on identifying a spin-split band structure. 
However, studying the spin-splitting from the band structure is not a useful signature for surface AM. The characteristic altermagnetic spin density can emerge as a local feature at the surface. Global bulk properties, e.g., the band structure, cannot necessarily detect this effect, because the contributions of opposing surfaces cancel out.

\begin{figure}
    \centering
    \includegraphics[width=0.9\linewidth]{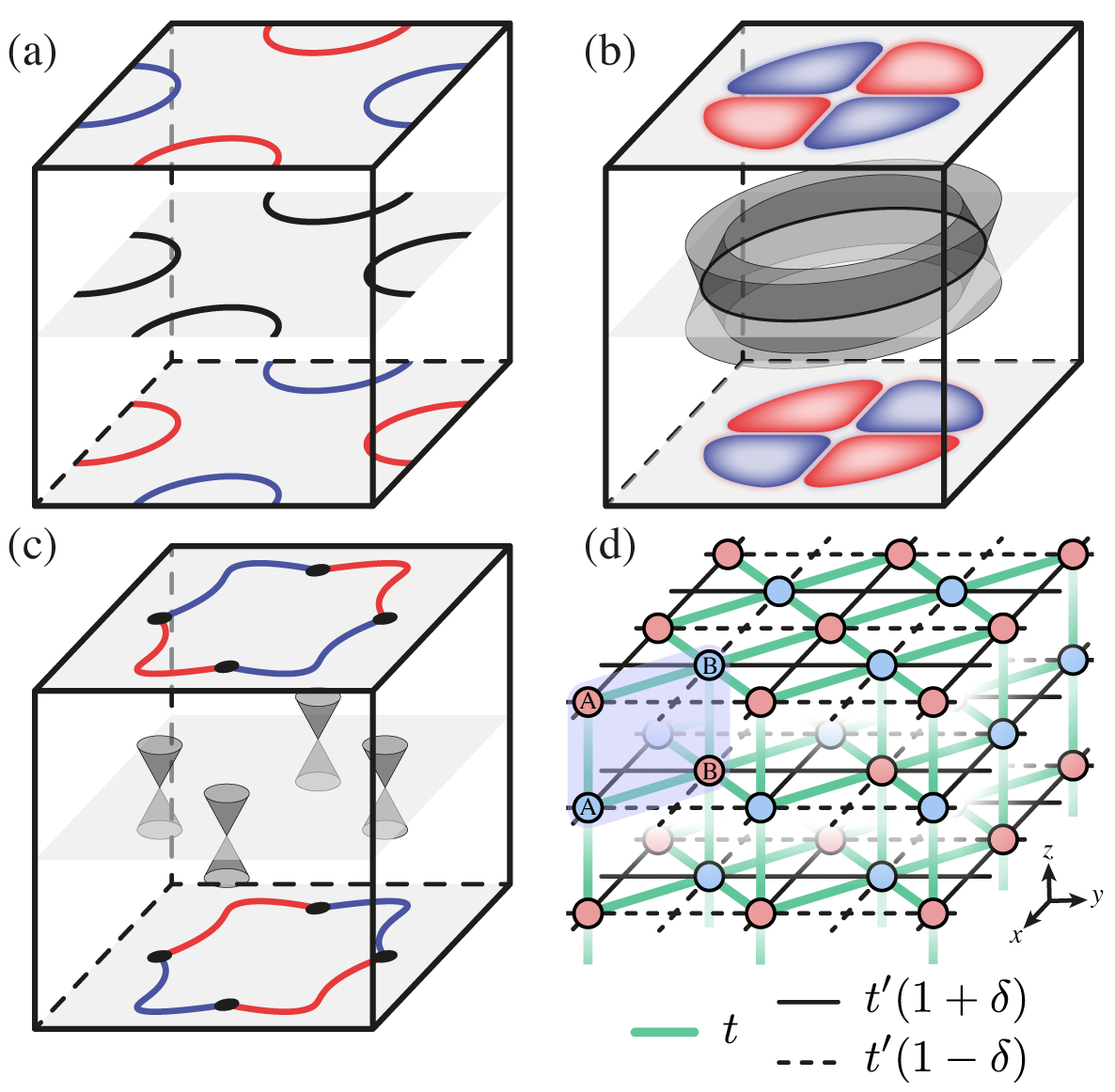}
    \caption{Altermagnetic surface states on spin-degenerate AFMs.
    (a) The antiferromagnetic bulk symmetry group can break into an altermagnetic surface symmetry group, leading to a spin-split surface Fermi surface (red,blue is the spin character) where the bulk is spin degenerate. (b) In AFM Dirac nodal line semimetals spin-degeneracy is lost at the surface resulting in topologically protected altermagnetic drumhead states. (c) In AFM Dirac semimetals altermagnetic Fermi arcs connect the bulk topological nodes.
    (d) To unearth the surface behavior described in (a-c) crystals are studied in a slab geometry, i.e., periodic boundary conditions along $x$ and $y$ direction and open boundary conditions in $z$ direction. Here, the crystal structure of \eqref{eq:H_DLKK} is shown, where spin-up (spin-down) sites are red (blue). The bulk 4-site unit cell is highlighted in blue, labeled by sublattice, and the layer degree of freedom is in the $z$ direction. 
   }
    \label{fig:fig_1}
\end{figure}

In this work, we demonstrate that spin-degenerate AFMs can exhibit altermagnetic spin-split characteristics at the surface. Such a system may be understood as a bulk AFM with local altermagnetic surface states.
Note this differs substantially from the surface states of bulk AMs \cite{sattigeri_altermagnetic_2023,gauswami_exploration_2025,fukaya_crossed_2025,sorn_antichiral_2025,li_topological_2025,lu_signature_2025,du_topological_2025,liu_tunable_2025,wan_helical_2025} or the influence of an altermagnetic layer on non-magnetic surface states \cite{jiang_altermagnetism_2025}.%

We report on both trivial and topologically-protected altermagnetic surface states, providing minimal models that realize both of them. First, we explain the general symmetry prerequisites for such surface states to emerge and demonstrate the existence of surface AM in a simple, topologically trivial system with spin-degenerate bulk bands and a spin-split surface, see \cref{fig:fig_1}~(a). Next, we use topology as a resource to ensure the altermagnetic surface is gapless and robust. 
Finding such a protected state is challenging because topology in collinear magnets is typically induced by spin-orbit coupling (SOC) \cite{Nagaosa10,Bernevig22}, however AM is not well-defined in the presence of SOC \cite{smejkal_emerging_2022}. Despite the symmetry conditions for band topology and surface AM being close to incompatible, we show that antiferromagnetic nodal line and Dirac semimetals can exhibit topologically protected altermagnetic surface states. We construct a realization of AM drumhead states, depicted in \cref{fig:fig_1}~(b), and AM Fermi arcs, depicted in \cref{fig:fig_1}~(c). Finally, we propose the room-temperature antiferromagnetic Dirac semimetals CuMnAs and CuMnP \cite{maca_roomtemperature_2012}, known for their applications in spintronics \cite{tang_dirac_2016,smejkal_topological_2018,smejkal_electric_2017}, as existing material realizations of our theory.

{\it Symmetry requirements---}%
Surface AM arises in bulk collinear AFMs when the non-altermagnetic bulk symmetry group breaks down at the surface to an altermagnetic symmetry group.
The symmetries that can enforce AM order at a surface are a combination of spin-flip times either a rotation symmetry (most commonly $C_{4z}$) or a mirror symmetry. In a \textit{crystalline context} this symmetry must be compatible with the 2D surface crystal structure itself, which restricts the candidate symmetries to mirrors and $C_{4z}$ by the crystallographic restriction theorem \footnote{The crystallographic theorem only admits $C_1$, $C_2$, $C_3$ $C_4$ and $C_6$. However, $C_2 \mathcal T$ implies spin degenerate bands and $C_6 \mathcal T$ has $(C_6 \mathcal T)^3 = C_2 \mathcal T$. $C_3 \mathcal T$ implies $C_1 \mathcal T$---i.e.~the system has time reversal symmetry. Thus, $C_4\mathcal T$ is the only rotation that can lead to altermagnetic bands. Note that here we restrict to the strictly 2D symmetries that emerge at a surface. In the bulk of a 3D material these restrictions no longer apply since symmetries can involve translation or reflection in the $z$ direction.}. Note that these symmetry constraints are more restricted than those for bulk AM in a 3D material, since the symmetry at the surface cannot involve any component (such as translation or reflection) in the direction normal to the surface, i.e.~the $z$ direction. Spin-group symmetries that survive at a surface have previously been shown to protect surface degeneracies and nodal structures~\cite{liu_spin-group_2022}. Here, we ask which surface symmetries permit spin splitting.
We start by providing the exact symmetry requirements for surface AM effect to occur. 

Without loss of generality we assume a (001) surface. Hence, there are two symmetries in a collinear magnet which protect spin-degeneracy in the bulk, but are absent at the surface: (i) an inversion ($\mathcal{I}$) times time reversal ($\mathcal T$) symmetry $[\mathcal{T} \lline \mathcal{IT}]$ and (ii) a $C_2$ rotation in spin space, effectively acting as a spin inversion, times a fractional, (intra-unitcell) translation that involves some translation $\vc{t}_z$ in the $z$-direction $[C_2 \lline\vc{t}_z \vc{t}_\text{any}]$---here $\vc{t}_\text{any}$ indicates that the translation can have any component in the $x$ or $y$ direction.
%
%
At least one of these two symmetries has to be present to get a bulk AFM, but there must not be any other symmetry that protects spin degeneracy and is robust at the surface, e.g., $[C_2 \lline\vc{t}_\text{x or y}]$. Additionally, there must not be any $[\mathcal{T} \lline C_{2z}\mathcal{T}]$ axis, which would forbid spin splitting at the (001) surface. Finally, one combined spatial and time-reversal symmetry is required to enforce a vanishing magnetic moment at the surface, e.g. $[\mathcal{T} \lline \mathcal{T} C_{4z}]$ and $[\mathcal{T} \lline \mathcal{T} M]$ where $M$ is a mirror, where the mirror can also include translation---producing a glide symmetry. If all these requirements are met, the surface state is conventionally altermagnetic \footnote{Exceptions occur when the non-magnetic (e.g. orbital) part of the state breaks TRS which can lead to odd-wave magnets and compensated ferrimagnets~\cite{leeb_collinear_2026}.}.

{\it Symmetry-enforced AM---}%
We start by providing a simple 3D model of a bulk AFM with a \emph{spin-degenerate} band structure that features an altermagnetic spectral density at the surface. We consider a tetragonal 3D lattice, shown in \cref{fig:fig_1} (d). Since the lattice is bipartite, a N\'{e}el state where the nearest neighbor of a spin-up site is spin-down and vice versa is the conventional magnetic instability. Physically, such a magnetic state could emerge from a number of mechanisms, for example induced by a super-exchange mechanism in a Hubbard model with dominant nearest-neighbour couplings around half filling, see supplementary material (SM). Additionally, we include sub-leading, anisotropic next-nearest-neighbor (NNN) hoppings within each $xy$ plane, which induce a hopping between sites of the same sublattice. 

Hence, we consider the following Hamiltonian, 
\begin{align}
\begin{aligned}
    H_\text{triv} =
    &+ m\sum_{i,s} (-1)^{i+s} c_{i,s}^\dagger c_{i,s}
    -t\sum_{\langle i j \rangle, s} c_{i,s}^\dagger c_{j,s} 
    \\
    &-t'\sum_{\langle\langle i j \rangle\rangle \perp \hat e_z, s}(1+(-1)^i \delta) c_{i,s}^\dagger c_{j,s},
    \label{eq:H_DLKK}
\end{aligned}
\end{align}
where $m$ is the local magnetization that leads to the formation of N\'{e}el ordering, $s\in \{0,1\}$ labels the spin-down and spin-up states of a spin $\frac 12$ degree of freedom, $t=1$ parametrises the nearest neighbour hoppings, $t'$ controls the strength of the next-nearest neighbour hoppings within each layer, and $\delta$ determines the degree of anisotropy in the NNN hoppings, see \cref{fig:fig_1} (d). Here, we consider only this non-interacting effective Hamiltonian, however \eqref{eq:H_DLKK} may be straightforwardly derived as mean-field description of a Hubbard Hamiltonian with sub-leading NNN couplings, see SM. 

The magnetic state is a $(0,0,\pi)$ AFM with a four-site unit cell. The four sites of the magnetic unit cell $(0,0,0)$ (sublattice A, even layer), $(1/2,1/2,0)$ (sublattice B, even layer), $(0,0,1/2)$ (sublattice A, odd layer), and $(1/2,1/2,1/2)$ (sublattice B, odd layer) can be grouped into sublattice and layer, highlighted in blue in \cref{fig:fig_1} (d). The two sublattices within each layer are crystallographically inequivalent, but related by a $\pi/2$-rotation, and have opposite magnetization. The layers are crystallographically equivalent, but have inverted N\'{e}el vectors.

\begin{figure}
    \centering
    \includegraphics[width=\columnwidth]{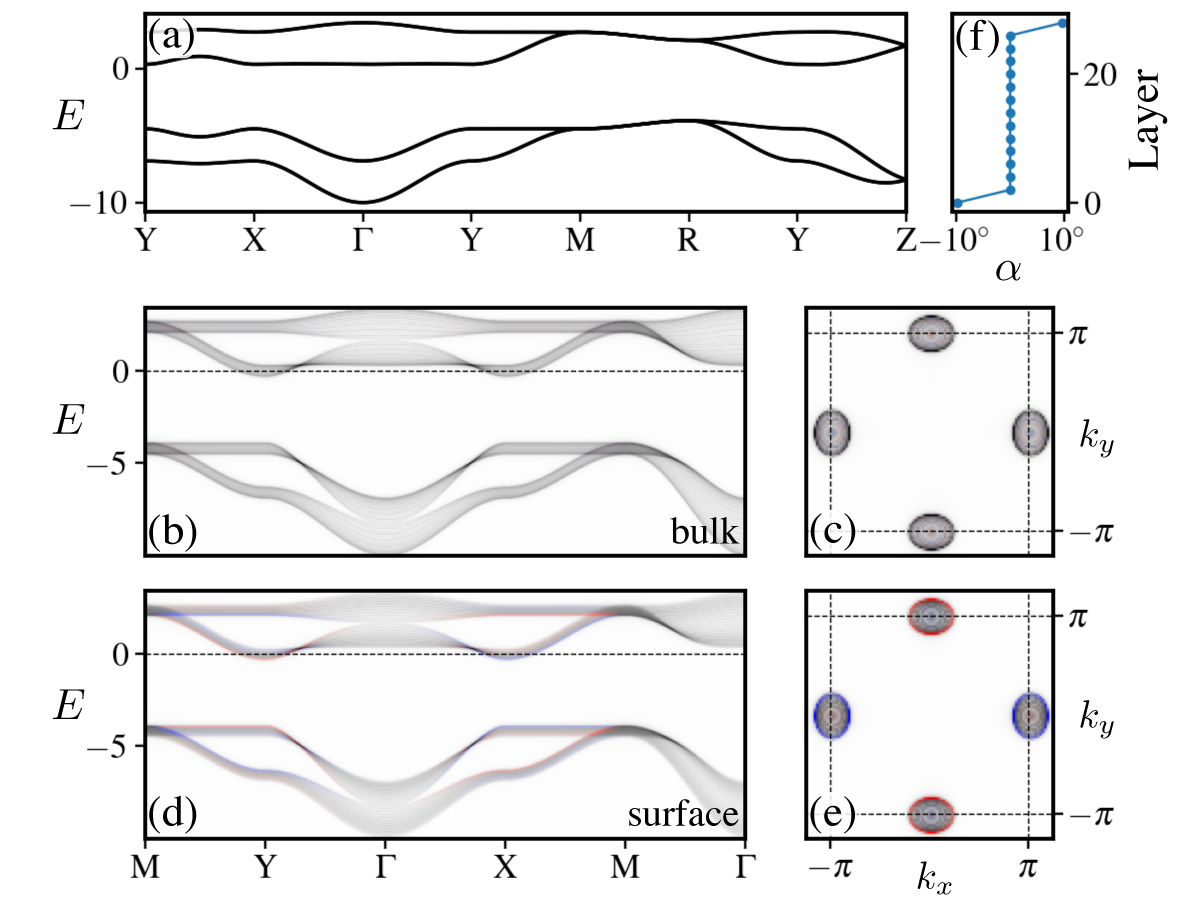}
    \caption{
    Topologically trivial altermagnetic surface state in a spin degenerate AFM, defined in \cref{eq:H_DLKK}. 
    (a) The 3D bulk band structure is spin-degenerate.
    (b-e) The model in slab geometry, with $L_z = 30$.
    Colorplots show the spin-resolved spectral function projected onto a pair of layers. We use the following color scheme: 
    $\mathcal{A}_{\ua} - \mathcal{A}_{\da}$ sets the coloring from spin-down polarized (blue) over spin-degenerate (black) to spin-up polarized (red); $\mathcal{A}_{\ua} + \mathcal{A}_{\da}$ sets the opacity. 
    (b,c) The spin spectral density $\mathcal{A}_{s}(z=L_z/2)$ projected to the bulk along a high symmetry path (c) and at the Fermi energy ($E=0$, dashed line) (d).
    (d,e) The spin spectral density $\mathcal{A}_{s}(z=0)$ projected at the surface along a high symmetry path (c) and at the Fermi energy (d).
    (f) The layer-resolved spin conductivity, expressed as the spin splitter angle $\alpha = 2\arctan((\sigma^\ua_{xx}-\sigma^\da_{xx})/(\sigma^\ua_{xx}+\sigma^\da_{xx}))$ shows that the the spin splitting is only sizable at the surface. 
    }
    \label{fig:DLKK_model}
\end{figure}

Individually, each $xy$-layer of this model would be $d$-wave altermagnetic, because its magnetic sublattices are related by a $\pi/2$-rotation; to this checkerboard square lattice is referred as DLKK model \cite{das_realizing_2024,langmann_mixed_2025, langmann_update_2025} or $t$-$t'$-$\delta$-Hubbard model
\cite{das_realizing_2024,he_altermagnetism_2025}.
However, neighboring layers have inverted spin patterns, ensuring that the spin splitting on odd layers perfectly compensates that on even layers. Thus, the overall magnetic state forms a bulk AFM. 
Formally, the spin up and spin down sublattices are related by a fractional translation $\vc{t}_z$, and inversion $\mathcal{I}$, which maps adjacent layers onto one another. In spin-space group notation the symmetries $[\mathcal{T}\lline\mathcal{T}\mathcal{I}]$ and $[C_2\lline\vc{t}_z]$ protect the spin degeneracy of the band structure, see Fig.~\ref{fig:DLKK_model}~(a).

At a (001) surface the $[\mathcal{T}\lline\mathcal{T}\mathcal{I}]$ and $[C_2\lline\vc{t}_z]$ symmetry are broken. Hence, the surface states in this antiferromagnet are not locally protected against spin splitting. To see this, we consider \cref{eq:H_DLKK} in a slab geometry, i.e.,~a lattice of $L_z$ $x$-$y$-layers with open boundary conditions in the $z$-direction and periodic boundary conditions in the $x$- and $y$-directions. Retaining $x$-$y$ translation symmetry allows us to transform to momentum space $\vc{k}=(k_x, k_y)$, such that the $L_z$ layers appear as $4 L_z$ different bands in the slab-geometry band structure, see Fig.~\ref{fig:DLKK_model}~(b). This model is physically motivated from a Hubbard model---from unrestricted mean-field simulations we found that the magnetization in the 3D slab system is roughly constant throughout the system, see SM.

Surprisingly, even the slab band structure is spin degenerate. This is due to a global inversion times time-reversal symmetry $[\mathcal{T}\lline \mathcal{T}\mathcal{I}]$, which forces momentum states with opposite spin at opposite surfaces of the system to have the same energy, see SM \footnote{~In fact the global inversion symmetry only exists if there is an even number of layers. For an odd number of layers it is weakly broken but this is an $\frac{1}{L_z}$ effect and means that the dispersion of the surface states on opposite sides of the system is slightly different. We expect this splitting in general to be weakly present, but is not the defining spin splitting. This \emph{is} the splitting which has been identified as surface AM in previous works \cite{mazin_induced_2023,zhou_transition_2025,meier_antialtermagnetism_2025}.}.
However, this has deeper implications: Even though spin up and down $\vc{k}$-momentum states are symmetry-constrained to have the same energy, the symmetry constraint is highly non-local. It relates opposite sides of the system, but each individual surface is not protected against spin splitting. Hence, we can observe a spin-split surface state on one side of the system and its spin-inverted counterpart on the other side of the system. 

We visualize the spin-, momentum-, and space-resolved spectral density by a spin-resolved spectral function, which may be written in terms of the Greens function, $G(\omega, \vc{k})$ as 
\begin{equation}
\mathcal A_s(\omega,\vc{k},z) = -\frac 1 \pi \Im \tr \left [(P_{z} + P_{z+1}) G_s(\omega, \vc{k})\right ],
\label{eq:spectral function}
\end{equation}
where $P_z$ is a projector onto the layer at $z$. \cref{fig:DLKK_model}~(c,d) shows that deep within the bulk the spectral density is spin-degenerate because inversion and $z$-translation are locally approximately conserved. However at the surface, see \cref{fig:DLKK_model} (d,e), the surface spectral density is partly $d$-wave spin polarized.

This is \emph{not} an ultra-high resolution single layer effect \cite{meier_antialtermagnetism_2025}, but a consistent surface and subsurface spin splitting. In \eqref{eq:spectral function} it is therefore essential to study pairs of layers, because it is the bulk magnetic unit cell.

Finally, we also computed the layer-resolved (again we consider pairs of layers) spin conductance, see SM for details. Here, we use the spin-splitter angle $\alpha$ to quantify the spin-splitting, i.e., the angle between the spin-up and down current which is determined via the ratio of spin over electric conductivity~\cite{gonzalez-hernandez_efficient_2021}. As expected from the spin-split surface state, the spin conductance is only non-zero at the surface and odd under global inversion, see Fig.~\ref{fig:DLKK_model}~(f).

\begin{figure*}
    \centering
    \includegraphics[width = \textwidth]{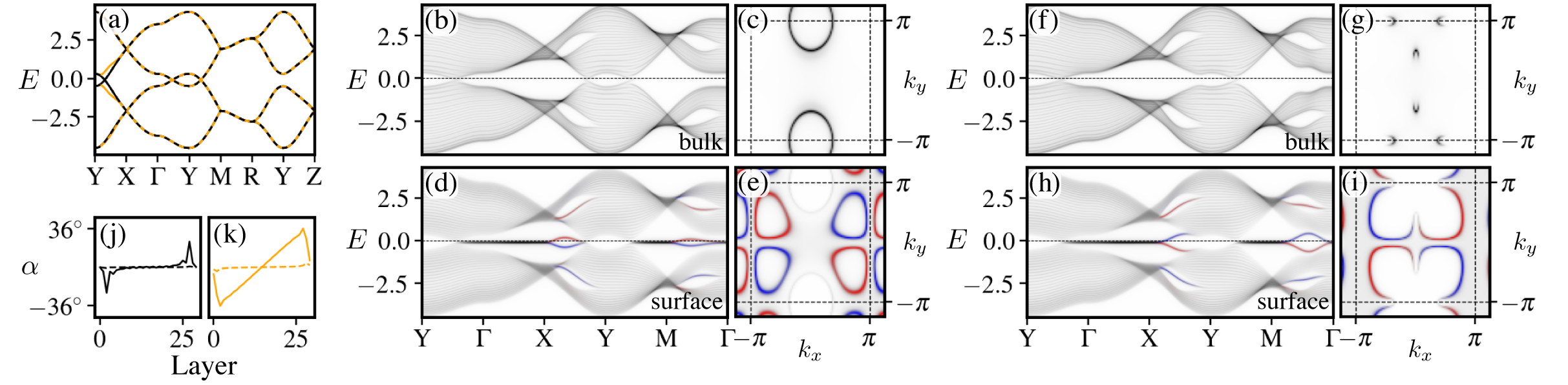}
    \caption{
    Topologically protected altermagnetic surface states in an AFM, described by $H_\text{topo}$ following \cref{eqn:h0,eq:H CuMnAs magnetic hoppings}. 
    (a) The 3D bulk band structure of the Dirac nodal line semimetal (black, $t'=0$) and the Dirac semimetal (yellow, $t'=0.2$) is spin degenerate. Dashed lines indicate overlap of both band structures. The touching points at $E=-0.1$ (shifted with a small chemical potential $\mu$) are the nodal line and the Dirac points.
    The spin-resolved spectral function in a slab geometry ($L_z = 31$) for the Dirac nodal line semimetal is shown in (b-e) and the Dirac semimetal in (f-i). The color scheme is identical to Fig.~\ref{fig:DLKK_model}.
    Panel (b,f) and (d,h) show the bulk and surface spin spectral density of each model along a high symmetry path respectively.
    This is projected deep in the bulk ($z=L_z/2$) for panels (b,f) and projected to the surface ($z=0$) for panels (d,h).
    The bulk spin spectral density $\mathcal{A}_s(z=L_z/2)$ at the Fermi energy ($E=0$, dashed line) shows the Dirac nodal line (c) and the Dirac cones (g). The surface spin spectral density $\mathcal{A}_s(z=0)$ at the Fermi energy shows the corresponding surface states, where closed contours are crossings of the drumhead states and the Fermi energy (e) and open arcs are crossings of the Fermi arcs and the Fermi energy (i).
    (j,k) The layer-resolved spin-splitter angles $\alpha_x$ ($\alpha_y$), as solid (dashed) line, which can be observed when applying an electric field in $x$-direction ($y$-direction). Panels (j) and (k)  correspond to the Dirac nodal line (black) and the Dirac semimetals (orange), respectively. 
    }
    \label{fig:drumhead_model}
\end{figure*}%

{\it Topological protection in CuMnAs---}%
The altermagnetic surface states discussed above are not topologically protected and  therefore might be hard to observe in practice. Imperfect surface termination and disorder can gap them out. In this section, we explain how topologically protected altermagnetic surface states emerge in spin degenerate AFMs and study the effect of surface disorder in the SM. We construct a minimal model that can be tuned to feature altermagnetic drumhead states and Fermi arcs. This minimal model is inspired by the room temperature AFM CuMnAs in its orthorhombic polymorph, reproducing their symmetry properties and band structure features, as we derive rigorously in the SM. 

Our starting point is a collinear, inversion odd magnetic order. For simplicity, we consider the same lattice structure and magnetic unit cell as above, see \cref{fig:fig_1} (d), where the atomic sites represent the 4a Wyckoff position of orthorhombic space group Pnma (SG \#62). Hence, the momentum-space Hamiltonian $H_\text{topo} = \sum_{\vc{k}} \vc{\Psi}_{\vc{k}}^\dagger h(\vc{k}) \vc{\Psi}_{\vc{k}}$ and its $8\times 8$ Bloch matrix $h$ are composed of the sublattice space $\sigma$, the layer subspace $\tau$, and spin space $s$. We consider two contributions to the Bloch matrix, $h= h_0+h_M$. The non-magnetic part of the Hamiltonian,
\begin{equation} \label{eqn:h0}
    h_0 = 2t \sigma_x \left(\cos \tfrac{k_x +k_y}{2}+\cos \tfrac{k_x -k_y}{2}\right) + 2t_\perp \tau_x \cos \tfrac{k_z}{2},
\end{equation}
includes only in-plane $t$ and out-of-plane $t_\perp$ NN hoppings. 

We consider a magnetic order with the same symmetry as found in CuMnAs, which is inversion odd and hence preserves $[\mathcal{T}\lline\mathcal{T I}]$. In addition, the magnetic order also preserves a glide mirror symmetry combined with time reversal symmetry $[\mathcal{T}\lline\mathcal{T M}_x]$. 
Since the sites are the inversion centers, the magnetic order is described by spin-dependent hoppings instead of spin-dependent onsite potentials. Note that these hoppings are not SOC terms; unlike SOC they are time reversal odd. 
Considering only up to NNN magnetism-induced hoppings, the magnetic component of $h$, $h_M$, has four symmetry-allowed spin-dependent terms
\begin{align}
    h_M =& +2\Delta_0 \sigma_y s_z \left(\sin \tfrac{k_x +k_y}{2}-\sin \tfrac{k_x -k_y}{2}\right)
    \nonumber \\
    &+ 2\Delta_1 \sigma_z \tau_y s_z \sin \tfrac{k_z}{2}
    \nonumber \\
    &+2\Delta_2 \sigma_y \tau_x s_z \left(\sin \tfrac{k_x +k_y}{2}-\sin \tfrac{k_x -k_y}{2}\right)\cos \tfrac{k_z }{2}
    \nonumber \\
    &+ 2\Delta_3 \sigma_x \tau_y s_z \left(\cos \tfrac{k_x +k_y}{2}-\cos \tfrac{k_x -k_y}{2}\right)  \sin \tfrac{k_z }{2} .
\label{eq:H CuMnAs magnetic hoppings}
\end{align}

In \cref{fig:drumhead_model} (a, b-e, j), we show an example system with parameters, $t = 0.4, t_{\perp}=1, \Delta_0 = \Delta_1=\Delta_3 = 0.6$. Additionally, we include a small chemical potential $\mu = 0.1$, shifting the Fermi level ($E=0$) slightly away from the middle of the spectrum. The 3D electronic band structure of the Hamiltonian $H_\text{topo}$ is spin-degenerate, see \cref{fig:drumhead_model} (a), due to the combined inversion time reversal symmetry $[\mathcal{T}\lline\mathcal{T I}]$.

Additionally, the band structure has a topological feature, a Dirac nodal line in the $k_z=0$ plane around the $Y$-point, which is visible as a feature in the bulk spectral density, see \cref{fig:drumhead_model} (b-c). This four-fold degenerate band crossing is induced by the magnetic hoppings $\Delta_0, \Delta_1$ and protected by a $z$-inverting mirror symmetry. Following the bulk-boundary correspondence, the Dirac nodal line implies the presence of a 2D surface state --- a so called drumhead state \cite{burkov_topological_2011}. This in-gap state is almost flat in energy, but becomes dispersive through the NNN magnetic hoppings $\Delta_3$ (or $\Delta_2$, we set $\Delta_2=0$), and is visible in the surface-projected spectral function, see \cref{fig:drumhead_model} (d-e).

Deep within the bulk, see \cref{fig:drumhead_model} (b-c), the spectral function is spin degenerate \footnote{~Here, the global inversion symmetry only exists exactly if there is an odd number of layers, because the inversion center is on the layer.}. However at the surface, see \cref{fig:drumhead_model} (d-e), the drumhead state is partly spin-polarized.
The spin polarization of the drumhead state on one surface is inverted with respect to the opposing surface. Each drumhead state is protected by weak topology and $d$-wave-like spin-split due to the $[\mathcal{T}\lline\mathcal{T M}_x]$ symmetry. The total spectral density at the Fermi level is dominated by the drumhead surface states. We observe that the altermagnetic spin texture leads to a large spin-split conductivity around each surface, see \cref{fig:drumhead_model} (j).

The Dirac nodal line is protected by a $z$-inverting mirror symmetry. By breaking this symmetry, e.g. by an $\varepsilon_{yz}$ shear strain, the topological phase changes from a Dirac nodal line semimetal to a Dirac semimetal \cite{maca_roomtemperature_2012,chan_Ca3P2_2016}. The space group is then $P2_1/c$ ($\#$14), with point group $2/m$.
In the following, we introduce a term which gaps out the nodal line into four gapless Dirac points, such that the surface state transforms into four altermagnetic Fermi arcs. This term is an intrasublattice, intralayer hopping along the (110) direction 
\begin{equation}
    H_{\textup{topo}} \rightarrow H_{\textup{topo}} + t' \sigma_z \tau_z \left[\cos (k_x +k_y)-\cos (k_x -k_y) \right],
\end{equation}
which is nonzero on the entire $k_z=0$ plane, except at the Brillouin zone boundary and along the $x$ and $y$-axes. 
Therefore, it gaps out the nodal line everywhere except at the four points where the nodal line crosses the $y$-axis or the BZ boundary. At these points, the system now has gapless Dirac points, which can be seen in the bulk spectral function, see \cref{fig:drumhead_model} (f-g).

The topological Dirac points enforce the formation of altermagnetic Fermi arcs between their projected locations on the surface \cite{kargarian_are_2016}. The Dirac points consist of a spin-up Weyl point and a spin-down Weyl-point of opposite charge. This implies that each Dirac point must have one spin-up and one spin-down Fermi arc connecting it to an inverted Dirac point. The combined $x$-mirror time reversal symmetry guarantees that the spin up and down Fermi arcs have a different $d$-wave-like momentum dependence, see \cref{fig:drumhead_model} (h-i), with the opposite surface having an inverted spin configuration.
The altermagnetic spin density dominates the density of states, such that the conductivity is strongly spin-split throughout the \emph{entire} system, see \cref{fig:drumhead_model} (k) \footnote{The spectral weight in the bulk of the Dirac AM is extremely low since the only gapless states appear around the Dirac points, allowing the edge contributions to dominate throughout the system.}.

Finally, we also derived the leading order symmetry-allowed SOC terms and studied their effect on the altermagnetic surface states, see SM. Certain SOC terms can lift the topological protection of the altermagnetic surface, however its characteristics remain persistent. Surprisingly, some manifestations of SOC seem to \emph{enhance} the altermagnetic character of the surface state, leading to a larger spin-split conductivity, see SM. In CuMnAs specifically, SOC is expected to be small, $\approx7$meV~\cite{emmanouilidou_magnetic_2017}.

{\it Conclusion---}%
We have proposed a physical mechanism for topologically protected surface altermagnetism---the formation of robust AM surface states in spin-degenerate AFMs. We first discussed the requirements to realize surface AM: the absence of any combined spin-inversion times in-plane translation symmetries, $[C_2\lline  \vc{t}_\text{x or y}]$, and out-of-plane time reversal two-fold rotation axis, $\mathcal{T} C_{2z}$. These guarantee that the spin-degeneracy is only protected by time-reversal times inversion symmetry, $\mathcal{I}\mathcal{T}$, or spin-inversion times out-of-plane translation symmetry, $[C_2\lline \vc{t}_z]$. Then, any spin compensated surface is altermagnetic. We illustrated a resulting trivial surface AM in a model where conventional Ne\'{e}l AFM induces a spin split surface. 

Symmetry requirements guarantee that surface states are altermagnetic. However, there is no assurance that they exist or are robust against perturbations. Without topological protection surface states can be destroyed by adding a symmetry-respecting term. We proposed two different mechanisms for topologically robust surface AM: (i) AM drumhead states in a Dirac nodal line semimetal and (ii) AM Fermi arc states in a Dirac semimetal. The bulk-boundary correspondence then guarantees topologically protected surface AM, as we demonstrate in the SM. 

The spin-splitting of altermagnetic surface states cannot be identified via global electronic or transport properties because opposite surface states compensate one another. Rather, the effect is only visible in the surface spin spectral density, which can be resolved by surface-focused,  spin- and angle resolved photoemission spectroscopy \cite{king_angle_2021, krempasky_altermagnetic_2024, lanzini_theory_2025,Krieger2024,Schrter2020}. Alternatively, we suggest that surface and spin-resolved transport measurements should be able to capture the local spin splitting~\cite{barreto_surfacedominated_2014,durand_differentiation_2016,hoefer_intrinsic_2014}.

Several existing materials fulfill the symmetry requirements to exhibit surface AM, and just need to be analyzed at the surface. In CuMnAs ab-initio calculation revealed a spin-split Fermi arc state under strain, in the related compound CuMnP it is yet to be confirmed \cite{maca_roomtemperature_2012}. Furthermore, $[C_2 \lline \vc{t}_z]$-protected AFMs consisting of spin-inverted AM layers \cite{meier_antialtermagnetism_2025}, e.g. induced by orbital ordering \cite{leeb_spontaneous_2024, meier_antialtermagnetism_2025,daghofer_altermagnetic_2025}, should show surface AM.
In general, materials where studies have focused on inducing AM by explicitly breaking inversion symmetry, e.g. through an electric field \cite{mazin_induced_2023,guo_externalfieldinduced_2025} are worth reanalyzing by considering the local surface physics. Photoemission experiments, discussed in the text, and quasi-particle interference, see e.g.~\cite{Inoue2016,wang_atomic_2025}, stand out as the most promising probes to distinguish topological surface states and their altermagnetic spin-textures.

Our work substantially enlarges the material landscape for AM and identifies topology as a resource for engineering robust spin-polarized electronic states at the surface of spin-degenerate AFMs. This is especially pertinent since a majority of magnetic materials are topological \cite{elcoro_magnetic_2021}, and AFMs are the most common form of magnetic ordering \cite{nmec_antiferromagnets_2018}. We therefore expect topological surface AM to provide a provide a new avenue of spintronics applications in AFMs and to provide a fertile ground for unconventional surface-driven phenomena. Finally, our work opens the possibility of extending this surface symmetry analysis to engineer a variety of novel collinear magnetic phases, such as collinear odd-wave magnets~\cite{leeb_collinear_2026} or type IV magnetism~\cite{bai_anomalous_2025}) on the surface of AFM and AM bulk materials.

{\it Data and code availability---}%
All code and data related to this paper, including the python package \texttt{blochK} \cite{blochk}, are publicly available \cite{alter_surf}.

{\it Acknowledgments---}%
We thank Q.~Meier for useful discussions. F.~J.~acknowledges funding from a 2024 Leonardo Grant for Scientific Research and Cultural Creation, BBVA Foundation. A.~G.~G.~and P.~D.~acknowledge support from the European Research Council (ERC) Consolidator grant under grant agreement No.~101042707 (TOPOMORPH).

{\it Note added---} Upon submitting this manuscript Refs.~\cite{Lange2026,Sasioglu2026} proposed surface altermagnetic surface states in antiferromagnetic materials without topological protection. During publication of this manuscript Ref.~\cite{yang_observation_2026} claimed to have observed surface altermagnetism in Cs$_{1-\delta}$V$_2$Te$_2$O, calling it `hidden altermagnetism'.



\bibliography{bib, references_manual} 

\clearpage
\onecolumngrid
\begin{center}
\textbf{\large Supplemental Material: Topologically Protected Surface Altermagnetism on Antiferromagnets}
\end{center}
\setcounter{equation}{0}
\setcounter{figure}{0}
\setcounter{table}{0}
\setcounter{page}{1}
\makeatletter
\renewcommand{\theequation}{S\arabic{equation}}
\renewcommand{\thefigure}{S\arabic{figure}}
\renewcommand{\bibnumfmt}[1]{[S#1]}
\setcounter{secnumdepth}{3}
\section{Spin conductance and spin splitter effect}
In our work we study a layer-resolved spin splitter effect, which is based on the the layer-resolved (spin) conductivity. Our calculation is based on the following setup: An uniform electric field is applied on the entire slab (parallel to the surface), the spin-resolved current of each pair of layers (one bulk unit cell) is measured. 

Here, we show how such a layer-resolved (spin) conductivity can be derived and computed. From a numerical perspective the slab Hamiltonian is a $4 L_z$ (2 sublattices per layer, 2 spins) dimensional matrix. By diagonalization we obtain $4 L_z$ Bloch bands $|u_{n}(\vc{k}=(k_x,k_y)) \rangle$ with dispersion $\epsilon_{n}(\vc{k})$. 

We define a conductivity tensor 
\begin{equation}
    \tilde\sigma_{\mu\nu}^\alpha(z) = \frac{e}{\pi V}\mathrm{Re} \sum_{\vec{k},m,n} \frac{\langle u_{n}(\vec{k})| J^\alpha_\mu(z) |u_{m}(\vec{k})\rangle \langle u_{m}(\vec{k})| v_\nu | u_{n}(\vec{k})\rangle \Gamma^2}{\left(\epsilon_{n}(\vec{k})^2 + \Gamma^2\right) \left(\epsilon_{m}(\vec{k})^2 + \Gamma^2\right)}
\end{equation}
where $v_c = \partial H(\vec{k})/\partial k_c$ is the velocity operator and $V$ the volume in analogy to previous works \cite{gonzalez-hernandez_efficient_2021, leeb_spontaneous_2024}. The current operator $J^\alpha_\mu(z)$ needs to capture both, spin and layer resolution. A spin current operator is typically defined as the symmetrized product of charge current (without an extra $e$) and spin operator $\{s^z, v_\mu\}/2$. In analogy, we define the layer-resolved (spin) current operator $J^0_\mu(z)$ ($J^z_\mu(z)$)
\begin{align}
    J^0_\mu(z) &= \{(P_z+P_{z+1}), v_\mu\} /2 \\
    J^z_\mu(z) &= \{s^z (P_z+P_{z+1}), v_\mu\} /2
\end{align}
where $P_z$ is the projector to a single layer. Note that $[P_z,s^z]=0$, but in general $[P_z,H]\neq 0$ because of next-nearest neighbor interlayer hoppings, i.e., hoppings with an $x$/$y$ and $z$ component. This definition ensures that the summation over all $z$ gives the conventional bulk conductivity. 

The spin-resolved conductivity $\sigma_{\mu\nu}^s(z)$ is simply a linear combination of the charge conductivity and the symmetric part of the spin-conductivity
\begin{align}
    \sigma_{\mu\nu}^\ua(z) &= \left(\tilde\sigma^0_{\mu\nu}(z) + \left[\tilde\sigma^z_{\mu\nu}(z)  + \tilde\sigma^z_{\nu \mu}(z) \right]/2 \right)/2
    \\
    \sigma_{\mu\nu}^\da(z) &= \left(\tilde\sigma^0_{\mu\nu}(z) - \left[\tilde\sigma^z_{\mu\nu}(z)  + \tilde\sigma^z_{\nu \mu}(z) \right]/2
    \right)/2.
\end{align}
The symmetrization of $\tilde\sigma^z_{\mu\nu}(z)$ is necessary because $[s^z (P_z+P_{z+1}), H] \neq 0$ can lead to an antisymmetric spin-Hall component. 

The spin-splitter effects occurs when the electric field is applied along the nodal direction. In the case of a $d$-wave like spin splitting there are 2 nodal directions, which are inequivalent in orthorhombic systems (like in our second model) and equivalent in tetragonal systems (like in our first model). Hence, there are 2 inequivalent spin splitter angles
\begin{align}
    \alpha_x = 2 \arctan \left( \frac{\sigma_{xy}^\ua-\sigma_{xy}^\da}{\sigma_{xx}^\ua+\sigma_{xx}^\da}\right) 
    \\
    \alpha_y = 2 \arctan \left( \frac{\sigma_{xy}^\ua-\sigma_{xy}^\da}{\sigma_{yy}^\ua+\sigma_{yy}^\da}\right) 
\end{align}
depending on if the field is applied in $x$ or $y$ direction. This applies for our second model, the minimal model for CuMnAs. 

In the first model, the $C_4$ symmetry enforces $\sigma_{xx} = \sigma_{yy}$ and the nodal lines and the axes of the coordinate system do not align. The spin splitter effect can still be evaluated by applying a principal axis transformation on $\sigma_{\mu\nu}$ which gives
\begin{equation}
    \alpha = 2 \arctan \left( \frac{\sigma_{xx}^\ua-\sigma_{xx}^\da}{\sigma_{xx}^\ua+\sigma_{xx}^\da}\right). 
\end{equation}

\section{Emergence of Spontaneous AM in a Hubbard Model} 
\label{apx:AM_hubbard}
\eqref{eq:H_DLKK} is an effective mean-field description of the electronic degrees of freedom with a magnetically ordered state in the background. Here, we show that \eqref{eq:H_DLKK} emerges microscopically from the Hubbard interaction. 

Consider the 3D Hubbard model
\begin{align}
    H_U =
    &
    -t\sum_{\langle i j \rangle, s} c_{i,s}^\dagger c_{j,s} 
    -t'\sum_{\langle\langle i j \rangle\rangle \perp \hat e_z, s}(1+(-1)^i \delta) c_{i,s}^\dagger c_{j,s}
    + U \sum_i c_{i,\uparrow}^\dagger c_{i,\uparrow} c_{i,\downarrow}^\dagger c_{i,\downarrow}
    \label{A:eq:HU_DLKK_}
\end{align}
whose hopping structure is identical to \eqref{eq:H_DLKK}. For dominant NN exchange, $t>t',\delta$ and around half-filling it is well understood that superexchange leads to an AFM state where nearest neighbor sites have opposite magnetization $\tilde m$. This motivates the mean-field ansatz 
\begin{equation}
\langle c_{i,s}^\dagger c_{i,s} \rangle = n/8 + (-1)^{i+s} \tilde m .  
\label{A:eq:DLKK MF decoupling 3D}
\end{equation}
The resulting magnetic state is a $(0,0,\pi)$-AFM, because there are already 2 sublattices in the $k_x$-$k_y$-plane due to the NNN hopping anisotropy $\delta$. Hence, the magnetic unit cell consists of 4 sites: 2 crystallographically identical layers (subspace $\tau$) times 2 sublattices (subspace $\sigma$) per layer. 

\subsection{MF decoupling in 3D}
Mean field decoupling the Hubbard interaction using \eqref{A:eq:DLKK MF decoupling 3D} and $A B \rightarrow \langle A \rangle B + A\langle B \rangle - \langle A \rangle \langle B \rangle$ gives
\begin{equation}
    U \sum_i c_{i,\uparrow}^\dagger c_{i,\uparrow} c_{i,\downarrow}^\dagger c_{i,\downarrow} 
    \rightarrow U \sum_{i,s} c_{i,s}^\dagger c_{i,s} \left(\frac{n}{8}-(-1)^{i+s}\tilde m\right) + 4 U N \left(\frac{n^2}{64} - \tilde m^2\right).
\end{equation}
We absorb the constant shifts into the chemical potential, renormalize the magnetization $m = U\tilde m $,  and write the momentum space Hamiltonian $H = \sum_{\vc{k}} \vc{\Psi}_{\vc{k}}^\dagger h(\vc{k}) \vc{\Psi}_{\vc{k}}$ in terms of a $8 \times 8 $ Bloch matrix, spanned by the sublattice ($\sigma$), layer ($\tau$), and spin ($s$) subspace.  The Bloch Hamiltonian
\begin{align}
    h(\vc{k}) =& 
     -2t \sigma_x \left[\cos k_x + \cos k_y \right] - 2t \tau_x \cos k_z 
     -2t' \left[\cos(k_x+k_y) + \cos(k_x-k_y)\right]
     \nonumber \\
    &-2\delta \sigma_z  \left[\cos(k_x+k_y) - \cos(k_x-k_y) \right] - \sigma_z s_z \tau_z  m
    \label{eq:h(k) DLKK}
\end{align}
is block-diagonal in spin, which leads to 4 spin-degenerate bands
\begin{align}
    E_{\pm,\pm,s}(\vc{k}) =& -2t' [\cos(k_x+k_y) + \cos(k_x-k_y)]
    \nonumber \\
    &\pm 
    \left[m^2 + 4t^2 \left(\cos k_x + \cos k_y \right)^2 + 4t^2 \cos^2 k_z + 4\delta^2 \left(\cos(k_x+k_y) - \cos(k_x-k_y) \right)^2 \right.
    \nonumber \\
    &\pm 
    \left.
 2 \left[(m^2 + 4t^2 \cos^2 k_z) (4t^2 \left(\cos k_x + \cos k_y \right)^2 + 4\delta^2 \left(\cos(k_x+k_y) - \cos(k_x-k_y) \right)^2)\right]^{1/2}
    \right]^{1/2}.
\end{align}
The spin-degeneracy is a result of the combined time reversal inversion symmetry $\mathcal{T I}$. In our model time reversal is $\ii s_y \mathcal{K}$ ($\mathcal{K}$ is complex conjugation) with $\vc{k} \rightarrow -\vc{k}$ and inversion is $\tau_x$ with $\vc{k} \rightarrow -\vc{k}$. Hence, the invariance can be explicitly observed by the fact that $h(\vc{k})$ is real and commutes with $s_y \tau_x$.

\subsection{MF decoupling of the slab}
Now we consider a slab, i.e., a finite amount of layers in $z$-direction. In the main manuscript we considered a constant density and magnetization over all layers. Here, we show that there is a layer dependence of density and magnetization, which does not affect our conclusions.

The surface determination motivates an ansatz which depends on the distance from surface determination. This means the mean-fields are $z$-dependent
\begin{equation}
\langle c_{(x,y,z),s}^\dagger c_{(x,y,z),s} \rangle = n_z/8 + (-1)^{x+y+z+s} \tilde m_z /2 .  
\label{A:eq:DLKK MF decoupling slab}
\end{equation}
We introduce the 4 component spinor $\vc{\Psi}_{k_x,k_y,z}$ which is only Fourier transformed in the $x$ and $y$ component. The mean field decoupling leads to 
\begin{equation}
    U \sum_i c_{i,\uparrow}^\dagger c_{i,\uparrow} c_{i,\downarrow}^\dagger c_{i,\downarrow} 
    \rightarrow U \sum_{k_x,k_y,z,s} \left(\frac{n_z}{8} \vc{\Psi}_{k_x,k_y,z}^\dagger \cdot \vc{\Psi}_{k_x,k_y,z} -(-1)^{z} \frac{\tilde m}{2} \vc{\Psi}_{k_x,k_y,z}^\dagger \tau_z \sigma_z \vc{\Psi}_{k_x,k_y,z} \right) - U L_x L_y \sum_z \left(\frac{n_z^2}{16} - \tilde m_z^2\right).
\end{equation}

The Hamiltonian
\begin{equation}
    H = U L_x L_y \sum_z \left(\tilde m_z^2 - \frac{n_z^2}{16}\right) + \sum_{k_x,k_y,z} \vc{\Psi}_{k_x,k_y,z}^\dagger h(k_x,k_y,z) \vc{\Psi}_{k_x,k_y,z} - t \vc{\Psi}_{k_x,k_y,z}^\dagger \cdot \vc{\Psi}_{k_x,k_y,z+1} + \text{h.c.}
    \label{A:eq:H_DLKK MF slab}
\end{equation}
then consists of the $z$-dependent Bloch matrix
\begin{align}
    h(k_x,k_y,z) = & 
    -2t \sigma_x \left[\cos k_x + \cos k_y \right] 
    -2t' \left[\cos(k_x+k_y) + \cos(k_x-k_y)\right]
     \nonumber \\
    &-2\delta \sigma_z  \left[\cos(k_x+k_y) - \cos(k_x-k_y) \right] + \frac{U}{8} n_z - \sigma_z s_z (-1)^z  U \tilde m_z
    \label{eq:h(k,z) DLKK}
\end{align}
and hopping elements, such that the Hilbert space is $4 L_z$ dimensional. 

\begin{figure}
    \centering
    \includegraphics[width=0.8\linewidth]{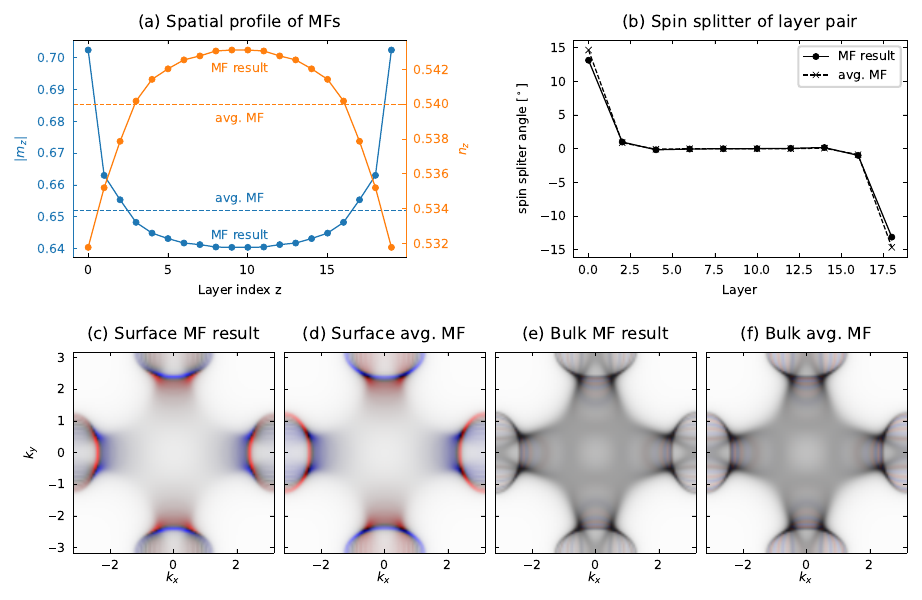}
    \caption{Real space mean field simulations of the slab confirm that assuming constant mean fields is a valid assumption. Mean field result of \eqref{A:eq:H_DLKK MF slab} for $t' = 0.3 t$, $U=6t$, $\delta=1$, around half-filling $n=0.54$. (a) The AFM magnetization $m_z$ tends to increase slightly towards the surface, whereas the density $n_z$ is almost constant. (b) Comparison of the spin splitter effects for layer-dependent mean field results (solid) and using the average values for the mean fields (dashed). (c-f) Comparison of the surface (c-d) and bulk (e-f) spin spectral density (opacity encodes absolute spectral density and the red to blue encodes the spin character) when using the layer-dependent mean field values (c,e) and their mean values (d,f). }
    \label{A:fig:DLKK_MF_const}
\end{figure}

We performed combined restricted (in $x$ and $y$ direction) and unrestricted (in $z$ direction) Hartree--Fock mean field calculations. We compared the energy of the solution of the ansatz above to non-magnetic and ferromagnetic solutions. For sufficiently large $U$ and close to half-filling the $(0,0,\pi)$ AFM with a weak $z$-dependence, see Fig.~\ref{A:fig:DLKK_MF_const}, is consistently the lowest energetic state.

We solve the mean-field equation iteratively using a $\vc{k}$-mesh of size $50 \times 50$ and 20 layers, resulting into 40 independent mean fields. The combined real-space ($z$-direction) and momentum-space ($x$ and $y$-direction) mean field simulations on the slab confirm that assuming constant mean fields is a valid assumption. The phenomenology we described does not depend on the surface dependence of any values but simply on the fact that there exists a surface. 

\section{Minimal model for orthorhombic \NoCaseChange{CuMnAs}}

In this section, we perform a symmetry analysis of CuMnAs and construct the minimal tight-binding model exhibiting altermagnetic, topologically protected surface states. 

The bulk polymorph of CuMnAs has orthorhombic space group Pnma (SG \#62). Note a tetragonal structure with space group P4/nmm (SG \#129) also exists, but is not the focus of this work \cite{wadley_tetragonal_2013}. Pnma is a non-symmorphic space group generated by three mutually orthogonal mirror symmetries $M_a\{\tfrac{1}{2},\tfrac{1}{2},\tfrac{1}{2}\}$, $M_b\{0,\tfrac{1}{2},0\}$ and $M_c\{\tfrac{1}{2},0,\tfrac{1}{2}\}$. We take the coordinate system such that $(y,z,x)$ correspond to $(a,b,c)$ axes. All three atoms (Cu,Mn,As) sit on 4c positions \cite{emmanouilidou_magnetic_2017}. Magnetic moments are located at the Mn atoms in an AFM pattern which preserves $\mathcal{IT}$ symmetry. In addition, in the absence of SOC the symmetry of the magnetic order can be analyzed separately in spin and spatial sectors, and the magnetic order is odd under the mirror $M_c$ but even under $M_a$ and $M_b$. At the surface normal to the $b$ axis, $\mathcal{IT}$ is broken but the combined symmetry $M_c \mathcal{T}$ is preserved in the absence of SOC, enforcing a zero total moment \cite{weber_local_2025} and thus qualifying it as an altermagnetic surface. 

In the absence of SOC, orthorhombic CuMnAs is a Dirac nodal line semimetal \cite{tang_dirac_2016,huyen_spin_2021,cao_inplane_2023} with drumhead surface states. When SOC is included, and if the N\'eel vector points along the $y$-direction, the nodal line is gapped out except at two Dirac nodes \cite{tang_dirac_2016,smejkal_electric_2017,kim_voltageinduced_2018}. Experimental evidence, however, points towards a N\'eel vector aligned along $z$ \cite{emmanouilidou_magnetic_2017,zhang_massive_2017} such that the nodal line is generically gapped by SOC. In this section, we first derive a model for the band structure without SOC.

\begin{figure}[t]
         \includegraphics[width=0.95\textwidth]{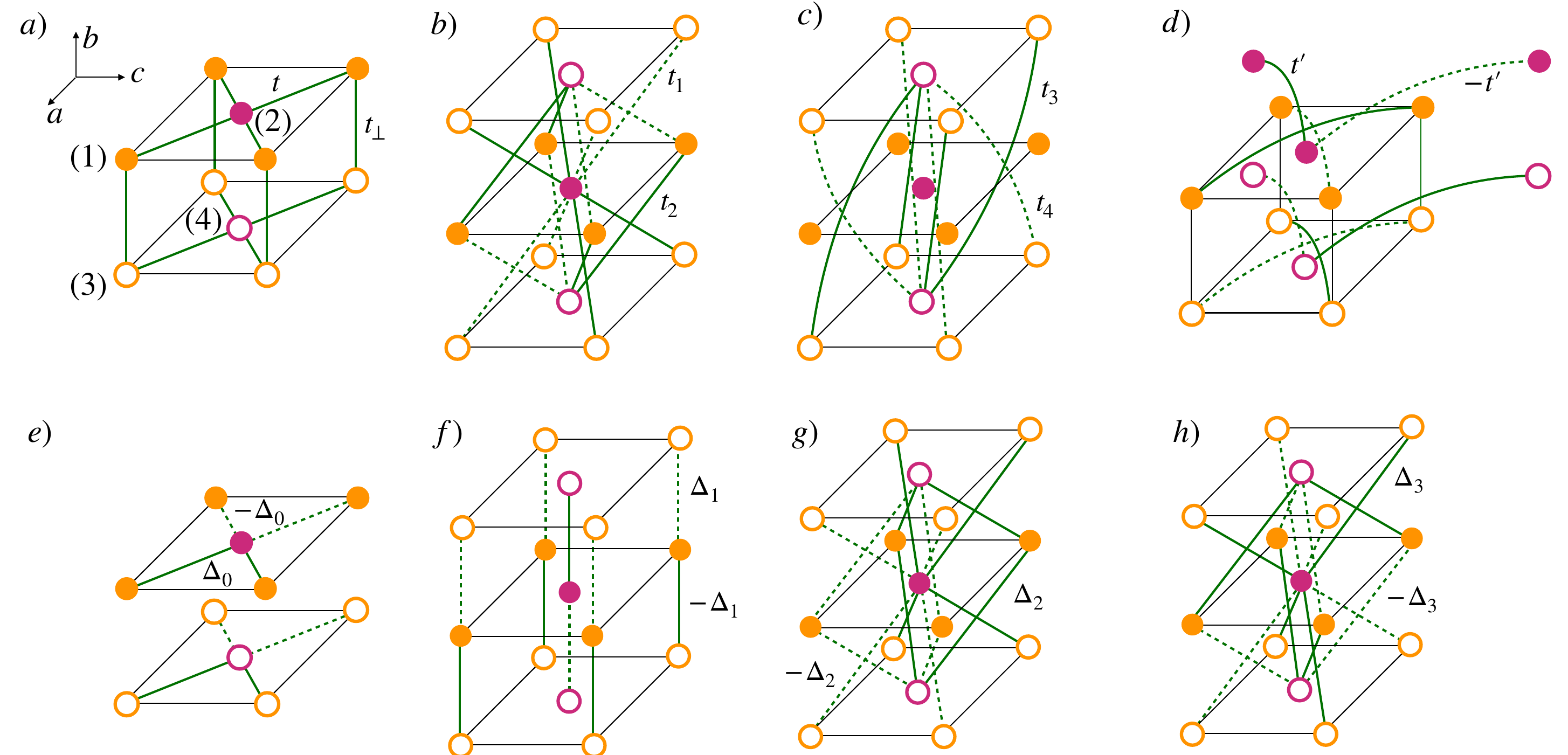}
     \caption{Tight-binding model for Wyckoff position 4a of Pmna. Filled (empty circles) are sites on even (odd) layers accounted by $\tau$-Pauli matrices, yellow (purple) circles are sites of the A (B) sublattices accounted by $\sigma$-Pauli matrices (a) In-plane and out-of-plane hoppings which preserve all three accidental translations in Eqs.~(\ref{A:eq:cumnas:translation1})-(\ref{A:eq:cumnas:translation3}). (b) Further neighbor hoppings which keep only $\vc{t}_{a/2,b/2,c/2}$. (c) Further neighbor hoppings which keep only $\vc{t}_{a/2,c/2}$. (d) (110) hopping $t'$, which gaps out the nodal line into 4 Dirac cones. 
     (e-h) Spin-dependent hoppings corresponding to the magnetic order with odd $M_c\{\tfrac{1}{2},0,\tfrac{1}{2}\}$ symmetry. }
     \label{A:fig:CuMnAs_structure}
\end{figure}

The nodal line occurs at $k_z=0$, and is protected by the mirror $M_b$ which leaves the $k_z=0$ plane invariant. The nodal line occurs within a subset of four bands, which come in pairs of opposite $M_b$ eigenvalues. Since Mn sits in Wyckoff position 4c, which is invariant under $M_b$, a model with a single orbital at 4c necessarily has all four bands which are even or odd under $M_b$, which means it is impossible to build a four band model for the nodal line. This means the Wannier centre for such set of four bands must be in another position with multiplicity 4, which can be 4a or 4b. Both positions are not invariant under $M_b$ and either can be picked, so we choose 4a. The model is layered in the $b$ direction, with orbitals at $(0,0,0)$, $(\tfrac{1}{2},\tfrac{1}{2},0)$ in the first layer and $(0,0,\tfrac{1}{2})$, $(\tfrac{1}{2},\tfrac{1}{2},\tfrac{1}{2})$ in the second layer. These sites are shown in Fig. \ref{A:fig:CuMnAs_structure}(a) and labelled (1-4) in this order. Taking the coordinate system where $(a,b,c) \leftrightarrow(k_y,k_z,k_x)$ and in the basis $(c_1,c_2,c_3,c_4)$ corresponding to the four sites in Fig. \ref{A:fig:CuMnAs_structure}(a), we define sublattice $\sigma_i$, layer $\tau_i$, and spin $s_i$ Pauli matrices.

\begin{align}
M_a: & &   H(k_x,k_y,k_z) &= \sigma_x \tau_x H(k_x,-k_y,k_z) \sigma_x \tau_x \\
M_b: & &   H(k_x,k_y,k_z) &=  \tau_x H(k_x,k_y,-k_z)  \tau_x \\
M_c: & &   H(k_x,k_y,k_z) &= \sigma_x H(-k_x,k_y,k_z) \sigma_x \label{eqn:mc} \\
\mathcal{I}: & &   H(\vec k) &= H(-\vec k)  &
\end{align}
The inversion center is at site 1, i.e., as expected at $(0,0,0)$.

\subsection{Non-magnetic part}
The simplest TB model has NN couplings in-plane $t$ and out of plane $t_\perp$
\begin{align}
H &= 2t \sigma_x (\cos \tfrac{k_x +k_y}{2}+\cos \tfrac{k_x -k_y}{2}) + 2t_\perp \tau_x \cos \tfrac{k_z}{2}  
\end{align}

The NN hoppings preserve spurious translation symmetries which are not part of the space group and can lead to spurious symmetries, which forbid AM splitting at the surface. In our example the spurious symmetries are broken by the magnetic order (the combination of the $\Delta_0$ and $\Delta_1$ term), therefore we do not need to consider the terms below. Nevertheless we discuss them for completeness.

The spurious translation symmetries are the three half-translations ($I$ is identity element)
\begin{align}
\vc{t}_{\tfrac{a}{2} \tfrac{c}{2}} &= I\{ \tfrac{1}{2}, 0 ,\tfrac{1}{2}\} &  H &= \sigma_x H \sigma_x 
\label{A:eq:cumnas:translation1} \\
\vc{t}_{\tfrac{b}{2} } &= I\{ 0, \tfrac{1}{2},0\} & H &=  \tau_x H \tau_x
\label{A:eq:cumnas:translation2} \\
\vc{t}_{\tfrac{a}{2} \tfrac{b}{2}\tfrac{c}{2}} &= I\{ \tfrac{1}{2},\tfrac{1}{2}, \tfrac{1}{2}\} &  H &= \sigma_x \tau_x H \sigma_x \tau_x 
\label{A:eq:cumnas:translation3}
\end{align}
which can give rise to accidental constraints for the band structure, as we show below. They can be broken by further neighbor hoppings, see Fig.~\ref{A:fig:CuMnAs_structure}(b). If we add
\begin{align}
     \delta H = &2 (t_1+t_2) \sigma_x \tau_x  \left[\cos \tfrac{k_x +k_y}{2}+\cos \tfrac{k_x -k_y}{2}\right]\cos \tfrac{k_z }{2}
+ 2 (t_1-t_2) \sigma_y \tau_y \left[ \sin \tfrac{k_x +k_y}{2}+\sin \tfrac{k_x -k_y}{2} \right] \sin \tfrac{k_z }{2}
\end{align}
with $t_1\neq t_2$, only $\vc{t}_{\tfrac{a}{2} \tfrac{b}{2}\tfrac{c}{2}}$ remains. If we add 
\begin{align}
\delta H &= 2\sigma_x [(t_3+t_4) \cos k_z (\cos \tfrac{k_x +k_y}{2}+\cos \tfrac{k_x -k_y}{2})]
+2\sigma_x \tau_z [(t_3-t_4) \sin k_z (\sin \tfrac{k_x +k_y}{2}-\sin \tfrac{k_x -k_y}{2})]
\end{align}
when $t_3 \neq t_4$, then only $\vc{t}_{\tfrac{a}{2} \tfrac{c}{2}}$ remains. If we include all then there are no spurious translations. 

The problem with spurious translations is that they may give rise to a spurious rotation symmetry $C_{2b}$. The actual group element of Pnma (SG \#62), $C_{2b}\{\tfrac{1}{2},\tfrac{1}{2},\tfrac{1}{2}\}$, is a combination of rotation and half translation along $b$ and is not preserved by the surface. However, combining  $C_{2b}\{\tfrac{1}{2},\tfrac{1}{2},\tfrac{1}{2}\}$ with the spurious translation symmetries gives
\begin{align}
 C_{2b}' = C_{2b}\{\tfrac{1}{2},\tfrac{1}{2},\tfrac{1}{2}\} *I \{0,\tfrac{1}{2},0\} 
 &= C_{2b}\{\tfrac{1}{2},0,\tfrac{1}{2}\} = \sigma_x\\
 C_{2b}'' = C_{2b}\{\tfrac{1}{2},\tfrac{1}{2},\tfrac{1}{2}\} * I\{ \tfrac{1}{2},\tfrac{1}{2}, \tfrac{1}{2}\}
 &= C_{2b} \{0,0,0\}
\end{align}
are both preserved by the surface. They enforce a spin degenerate surface state if only broken by the magnetic order, because then $[C_2\lline C_{2b}']$ or $[C_2\lline C_{2b}'']$ is preserved. 

\subsection{Magnetic part}
We assume a collinear magnetic order along the $z$ direction. The magnetic order is odd under the $x$-mirror $M_c$ and inversion $\mathcal{I}$. This implies that the magnetic order cannot be captured by onsite local potentials, because any site transforms under inversion back to itself and hence cannot be magnetic. 

Instead the simplest possible couplings with this symmetry are spin-dependent NN hoppings
\begin{align}
H_M =& 2\Delta_0 \sigma_y s_z(\sin \tfrac{k_x +k_y}{2}-\sin \tfrac{k_x -k_y}{2}) + 2\Delta_1 \sigma_z \tau_y s_z \sin \tfrac{k_z}{2}  
\end{align}
Both couplings are needed to generate the nodal line around $Y$ and to create a gap everywhere else. Note that these hoppings cannot be generated by SOC, which has different symmetry requirements (its spatial part is \textit{even} under inversion symmetry).  Together $\Delta_0$ and $\Delta_1$ break the spurious surface rotation symmetries: $\Delta_0$ is even under $C_{2b}'$ but odd under $C_{2b}''$, whereas the opposite is true for $\Delta_1$. The presence of both couplings is sufficient to destroy all $C_{2b}$ surface axes. 

However, the Hamiltonian with only $t$, $t_\perp$, $\Delta_0$ and $\Delta_1$ has a chiral symmetry
\begin{align}
H = -\chi H \chi    
\end{align}
with $\chi= \sigma_z \tau_z$, which enforces all drumhead states to have zero energy. The 4-fold degenerate bands associated with the surface states can hence not lead to any spin split spectral density at the surface. But again this symmetry is spurious and can be broken by further NN hoppings. Additional spin-dependent hoppings, respecting the symmetry of the magnetic order, are interlayer, intersublattice hoppings
\begin{align}
H_M'=
2\Delta_2 \sigma_y \tau_x s_z (\sin \tfrac{k_x +k_y}{2}-\sin \tfrac{k_x -k_y}{2})\cos \tfrac{k_z }{2}]
+ 2\Delta_3 \sigma_x \tau_y s_z(\cos \tfrac{k_x +k_y}{2}-\cos \tfrac{k_x -k_y}{2})\sin \tfrac{k_z }{2}]    
\end{align}
which break the chiral symmetry. In particular, $\Delta_3$ has an in-plane $d$-wave dependence and gives dispersion to the surface states.

\subsection{Symmetry breaking perturbations}

A symmetry breaking perturbation gaps out the Dirac nodal line. There are various way to do this. Here, we focus keeping spin conservation and gapping out the nodal line except for singular points --- these are then Dirac points. The surface of a spin conserved Dirac semimetal has Fermi arcs, because the topological charge of each spin species is conserved individually. A Dirac point consisting of a $C=+1$ spin-up Weyl and a $C=-1$ spin-down Weyl fermion must have both spin-up and a spin-down Fermi arcs that connect it to another Dirac point. 

The Dirac nodal line is protected by the $z$-inverting mirror $M_b$. 
In Ref.~\cite{tang_dirac_2016}, a perturbation that breaks $M_b$ but preserves $C_{2c} = M_b M_a$ and inversion was considered to gap out the nodal line into Dirac cones. This perturbation has the same symmetry as shear strain which preserves $M_c$, so $u_{ab}$ ($u_{yz}$). To reproduce this perturbation in our minimal model, we first consider all the possible constant on-site potentials with mirror parities $(M_a,M_b,M_c)$:
\begin{align}
    \delta H &= \mathcal{I} & & (+,+,+) \\
    \delta H &= \sigma_z & & (-,+,-) \\
    \delta H &= \tau_z & & (-,-,+) \\
    \delta H &= \sigma_z\tau_z & & (+,-,-)
\end{align}
Note because 4a is inversion symmetric, so are all these perturbations by construction. 

The perturbation $\tau_z$ therefore has the same symmetry as $u_{ab}$. However, it turns out that this perturbation is invariant under an accidental mirror $M_b'=\sigma_x$ of the $k_z=0$ plane, and the nodal line is not gapped unless higher order, symmetry-allowed hoppings are included. Because of this, we consider another perturbation with the same symmetry $\sigma_z \tau_z \left[\cos \left(k_x +k_y \right)-\cos \left( k_x -k_y \right) \right]$, which is the hopping considered in the main text, and shown in Fig.~\ref{A:fig:CuMnAs_structure}~(d).

\subsection{Spin-orbit coupling}

\begin{figure*}[t]
    \centering
    \includegraphics[width =\textwidth]{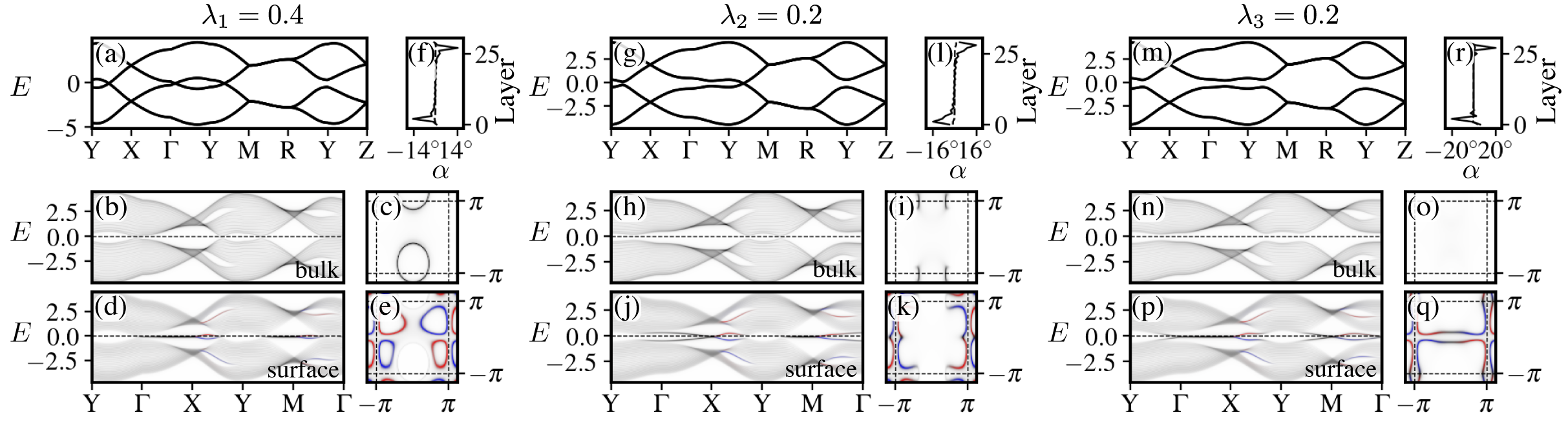}
    \caption{
    Influence of SOC on the altermagnetic drumhead states. We consider each of the three SOC terms in \cref{eqn:soc_terms}, each time setting the other two to zero. 
    (a-e) $\lambda_1=0.4$.
    Bulk 3D band structure (a), bulk spin-resolved spectral function (b,c), surface spin-resolved spectral function (d,e), and spin splitter angles $\alpha_x,\alpha_y$ (f). In the presence of $\lambda_1$-term SOC the spin-space group notation breaks down such that $[C_2||M_a]$ is not a symmetry anymore. 
    (g-k) $\lambda_2=0.2$.
    Bulk 3D band structure (g), bulk spin-resolved spectral function (h,i), surface spin-resolved spectral function (j,k), and spin splitter angles $\alpha_x,\alpha_y$ (l). 
    The $\lambda_2$-term SOC gaps out the Dirac nodal line everywhere except at the BZ boundary, where 2 Dirac cones persist and spin remains a good quantum number. 
    The topologically unprotected, altermagnetic surface state persists. 
    (m-q) $\lambda_3=0.2$.
    Bulk 3D band structure (m), bulk spin-resolved spectral function (n,o), surface spin-resolved spectral function (p,q), and spin splitter angles $\alpha_x,\alpha_y$ (r). 
    The $\lambda_3$-term SOC gaps out the Dirac nodal line everywhere.  The topologically unprotected, altermagnetic surface state persists. 
    }
    \label{A:fig:spin_orbit_drumhead}
\end{figure*}%

\begin{figure*}[t]
    \centering
    \includegraphics[width =\textwidth]{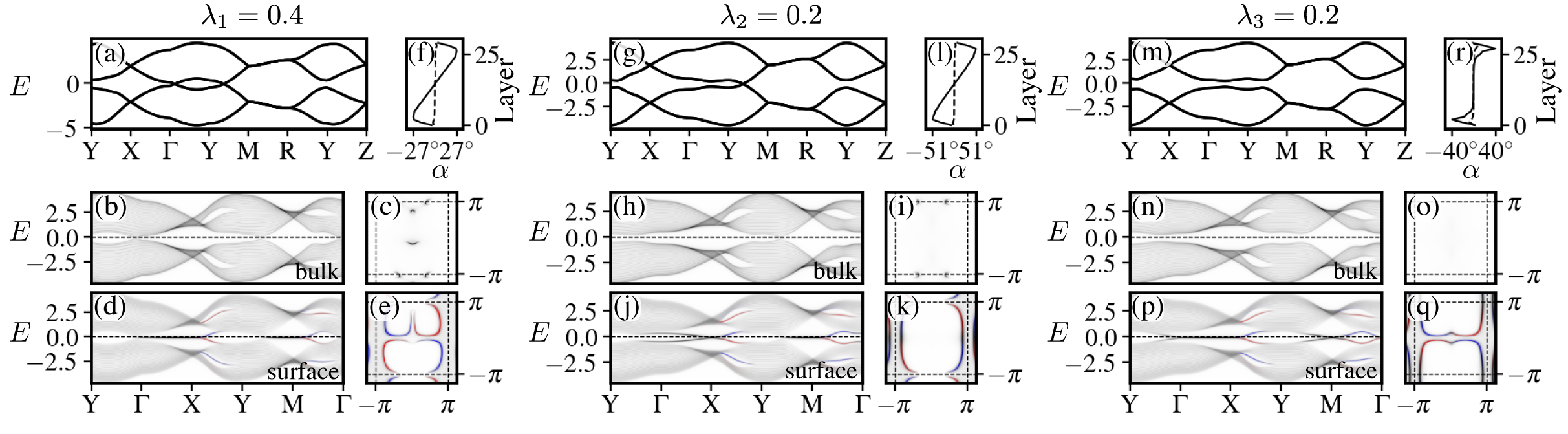}
    \caption{
    Influence of SOC on the altermagnetic Fermi-arc state. We consider each of the three SOC terms in \cref{eqn:soc_terms}, each time setting the other two to zero. 
    (a-e) $\lambda_1=0.4$.
    Bulk 3D band structure (a), bulk spin-resolved spectral function (b,c), surface spin-resolved spectral function (d,e), and spin splitter angles $\alpha_x,\alpha_y$ (f). In the presence of $\lambda_1$-term SOC the spin-space group notation breaks down such that $[C_2||M_a]$ is not a symmetry anymore. 
    (g-k) $\lambda_2=0.2$.
    Bulk 3D band structure (g), bulk spin-resolved spectral function (h,i), surface spin-resolved spectral function (j,k), and spin splitter angles $\alpha_x,\alpha_y$ (l). 
    The $\lambda_2$-term SOC gaps out the Dirac cones along the $y$-axis.The topologically unprotected, altermagnetic surface state persists.
    (m-q) $\lambda_3=0.2$.
    Bulk 3D band structure (m), bulk spin-resolved spectral function (n,o), surface spin-resolved spectral function (p,q), and spin splitter angles $\alpha_x,\alpha_y$ (r). 
    The $\lambda_3$-term SOC gaps out all Dirac cones.  
    }
    \label{A:fig:spin_orbit_Fermi}
\end{figure*}%

Since there is only one orbital per site, SOC is necessarily a hopping. Its spatial part must be time-odd, inversion even, and for spin matrix $s_i$ also odd under all mirrors except the $i$-th one. The lowest order hoppings that satisfy these requirements are 
\begin{align} \label{eqn:soc_terms}
 H_{\rm SOC} = \lambda_1 \sigma_y s_z (\cos \tfrac{k_x +k_y}{2}+\cos \tfrac{k_x -k_y}{2})
 + \lambda_2 \sigma_y \tau_z s_y (\cos \tfrac{k_x +k_y}{2}+\cos \tfrac{k_x -k_y}{2})
 + \lambda_3 \tau_y  s_x \cos \tfrac{k_z}{2}
\end{align}
which are intersublattice ($\lambda_1$), interlayer ($\lambda_3)$, and interlayer and -sublattice ($\lambda_2$) terms.

The SOC coupling terms have different effects on the electronic structure and the surface state, see Fig.~\ref{A:fig:spin_orbit_drumhead} and Fig.~\ref{A:fig:spin_orbit_Fermi}. In summary, we conclude that SOC does not affect the emergence of altermagnetic surface states. Instead, it can even significantly increase the spin splitter effect because the spin degenerate bulk density is gapped out. SOC can alter the topology of the band structure, but the altermagnetic character of the surface remains intact.

\subsection{Robustness of the surface state}

\begin{figure*}
    \centering
    \includegraphics[width = \textwidth]{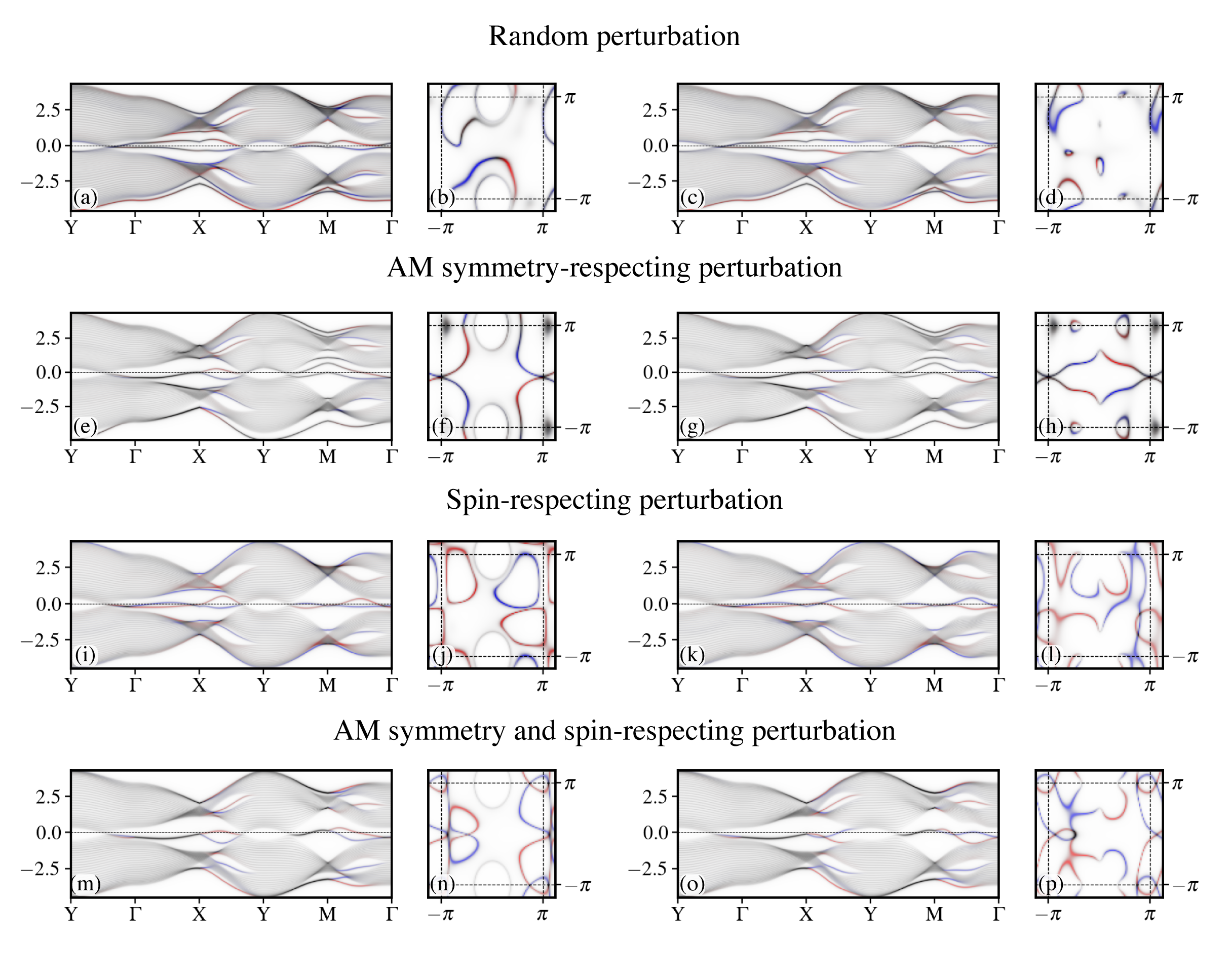}
    \caption{
    Four examples of the drumhead and Weyl system under the effect of a random surface perturbation. Each row corresponds to a different perturbation. The drumhead case is on the left and the Weyl case is on the right. The base models correspond to the two systems considered in \cref{fig:drumhead_model}, where here we only show the surface spectral functions, projected onto the top 3 layers (each layer containing 8 sites per unit cell)---note we use a few layers because perturbations can push the Fermi arcs a small distance into the bulk. For each case we show the spectral function along a path in the Brillouin zone, as well as a slice over the spectral function at the Fermi energy. 
    (a-d) A random perturbation is applied that breaks both spin and the altermagnetic $\mathcal M_c \mathcal T$ symmetry as well as spin conservation.
    (e-h) A random perturbation is applied that breaks spin symmetry but conserves the altermagnetic symmetry.
    (i-l) A random perturbation is added that preserves only spin symmetry but breaks the altermagnetic symmetry. 
    (m-p) A random perturbation is added that preserves both the altermagnetic and spin symmetry. 
    }
    \label{fig:perturbation}
\end{figure*}%

We now consider the robustness of the surface states that appear in the model both in the drumhead and Dirac cases. In the drumhead case, we expect that the existence of surface states is perfectly robust against perturbations, although it is possible to move the drumhead states in energy. The altermagnetic spin character of the drumhead states, and consequently the effect of surface altermagnetism should however be destroyed unless the perturbation itself respects the surface symmetry group. In the Dirac case, we should expect that generic disorder, which does not respect spin conservation, can gap out the Dirac nodes. This is because each Dirac node is composed of a pair of spin-up and spin-down Weyl nodes. Without spin-conservation it is possible to connect the spin-up and spin-down Fermi arcs to one another. Next, enforcing the altermagnetic symmetry will ensure that whatever surface states remain---Fermi arc or not---respect the altermagnetic surface symmetry, and so surface AM is preserved.

To this end, let us consider the effect of a randomly chosen surface perturbation on the surface states appearing in the two CuMnAs-based models presented in the bulk text. In this appendix, we only consider perturbations that respect translational symmetry, and so can be included as an arbitrary perturbation to the momentum-space Hamiltonian of the slab-geometry system,
\begin{align}
    H(k_x, k_y) \rightarrow H(k_x, k_y) + V(k_x, k_y),
\end{align}
where $H(k_x, k_y)$ is an $N\times N$ matrix with $N = 4L_z$ given by the height of the slab, and the unit cell contains sublattice, layer and spin degrees of freedom. Since we focus on surface effects, we only allow the top two layers of the material to be perturbed. Additionally, we restrict $V$ to only consider onsite and nearest-neighbor terms---of course, the same analysis can be extended to arbitrary-length hoppings, however the above choice is sufficient to illustrate the robustness of our surface states. Thus, the constraints in momentum space restrict us to the following five terms, 
\begin{align} \label{eqn:v_decomposition}
\begin{aligned}
    V(k_x, k_y) = &V_{0}^+ 
    + 
    V_{s}^+ \left (\sin \frac{k_x + k_y}{2} + \sin \frac{k_x - k_y}{2}\right )+ 
    V_{c}^+ \left (\cos \frac{k_x + k_y}{2} + \cos \frac{k_x - k_y}{2}\right )\\
    &+
    V_{s}^- \left (\sin \frac{k_x + k_y}{2} - \sin \frac{k_x - k_y}{2}\right )+ 
    V_{c}^- \left (\cos \frac{k_x + k_y}{2} - \cos \frac{k_x - k_y}{2}\right ).
\end{aligned}
\end{align}
The on-site component, $V_0$, does not hop between sublattices, so can only have terms proportional to $\sigma_0$ and $\sigma_z$, whereas the $V_s$ and $V_c$ terms exchange sublattice, so must be proportional to $\sigma_x$ and $\sigma_y$. Each component $V_{\mu}^\pm$, with $\mu \in \{0, s,c\}$, represents an $8\times 8$ matrix that is chosen from a uniform distribution of Hermitian matrices with real and complex component $\in [-1, 1]$. We consider three choices of $\{V^\pm_\mu \}$. First, we enforce no symmetries whatsoever on the perturbation, allowing a generic Hermitian $V$ for all components. 

Second, we enforce that the perturbation obeys the altermagnetic symmetry given by $[\mathcal{T}\lline\mathcal{T M}_x] $ in the language of the main paper (and spin space groups), or $M_c \mathcal T$ in the notation used in this appendix. Following \cref{eqn:mc}, we see that the effect of $M_c\mathcal T$ on the momentum space Hamiltonian is to apply the following transformation,
\begin{align} \label{eqn:mct_symmetry}
    M_c\mathcal T: \qquad H(k_x, k_y) \rightarrow \sigma_x s_y H^*(k_x, -k_y) \sigma_x s_y.
\end{align}
Inserting this into \cref{eqn:v_decomposition}, we see that the first three terms have a momentum component that is invariant under $k_y \rightarrow -k_y$, so the three matrices $V_\mu^+$ must satisfy 
\begin{align}
    \sigma_x s_y V_\mu^+\sigma_x s_y = V_\mu^{+*}.
\end{align}
Conversely, the last two terms pick up an overall sign under $k_y \rightarrow -k_y$, and so the two matrices $V_{\mu}^-$ must satisfy
\begin{align}
    \sigma_x s_y V_\mu^-\sigma_x s_y = -V_\mu^{-*}.
\end{align}
Each of these matrices is chosen at random, where we start with a completely general matrix, and remove the components that do not respect the desired symmetry to form the respective $V$ component. For the method, see the addendum below, \cref{enforcing_sym}. 

Next, we consider a perturbation that preserves the component of spin in the $z$-direction, but breaks the altermagnetic $\mathcal M_c \mathcal T$ symmetry. This is done by considering only components of $V^\pm_\mu $ which are proportional to either $s_0$ or $s_z$. Finally, we consider a perturbation that both respects the symmetry $\mathcal M_c \mathcal T$, \textit{as well as} preserving the component of spin in the $z$-direction.

The results are shown in \cref{fig:perturbation} for the four cases. In each, we use the same CuMnAs model that is studied in the main text, with the bulk and boundary spectral functions given in \cref{fig:drumhead_model}. We consider both the drumhead and Dirac system. Here we show the surface properties only, since the bulk of these models is identical to the bulk shown in \cref{fig:DLKK_model}. 

In the case of general disorder (shown in \cref{fig:perturbation}a-d), we see that surface altermagnetism is destroyed, since the perturbation is able to destroy the mirror symmetry that constrains the spectrum. In the drumhead case, the drumhead states are still present, however they have disordered spin character. In the Dirac case, we find two effects. Firstly, the altermagnetic symmetry is destroyed and so the spin character of the surface spectral function is effectively random. Secondly, since spin is not preserved, each Dirac node---which consists of a spin-up and spin-down Weyl node at the same point---still has a Fermi arc. However, the perturbation is able to connect the spin-up and spin-down node on the same side together, with a Fermi arc that exist with spin-up character, rotates in spin space and then re-enters the same point with spin-down character---which can effectively gap out the Dirac point on-site. 

In \cref{fig:perturbation}e-h we consider a perturbation that respects the altermagnetic surface symmetry but not spin conservation. We find that the surface altermagnetism is now robust. Here, the spin character of the spectral function in both the drumhead and Dirac cases is constrained such that the spin-up and spin-down components are related by a mirror over the $y$-axis. However, in the Dirac case it is still possible for for a Dirac node to connect to itself via a Fermi arc with changing spin character, and potentially gap out. 

Next, considering in \cref{fig:perturbation}i-l the case of a spin-preserving but $\mathcal M_c \mathcal T$-breaking perturbation, we see that the Drumhead states are separated into two drumheads, one spin-up and one spin down drumhead. Similarly, the Dirac case now has every Weyl node connected to a different node with the same spin orientation. However, the overall altermagnetic symmetry is lost. 

Finally, in \cref{fig:perturbation}m-p we show the case of disorder that respects \textit{both} the altermagnetic symmetry and preserves the component of spin in the $z$-direction, we find a robust altermagnetic state in which the drumhead states that emerge remain spin-split, with one drumhead state of each magnetization, and the Weyl nodes are completely intact, such that it is no longer possible for a Dirac node to connect to itself and gap out.  Furthermore, the drumhead and Weyl states are arranged such that the altermagnetic symmetry is respected.

\subsubsection{Addendum: Enforcing (Anti)Symmetry} \label{enforcing_sym}
Let us present the general problem first, given an idempotent matrix $\xi$ (in the case above we have that $\xi = \sigma_x s_y$) and a matrix $M$, we wish to find the component of $M$ that satisfies,
\begin{align} \label{eqn:ama}
    \xi M\xi = s M^*,
\end{align}
where $s \in \{\pm 1\}$ determines symmetry or anti-symmetry. The first step is to separate $M$ into a real and imaginary part $M = M_{\Re} + iM_{\Im}$. Inserting this into \cref{eqn:ama}, we find two conditions,
\begin{align}
    \xi M_{\Re}\xi  &= s M_{\Re},\\
    \xi M_{\Im}\xi  &= -s M_{\Im}.
\end{align}
Thus, we have reduced our problem to that of finding a pair of matrices, one which commutes with $\xi$ and one which anti-commutes with $\xi$. This can be done by starting with an arbitrary matrix $A$, and then taking the following two matrices, 
\begin{align} \label{eqn:Apm}
    A_\pm = \frac 12 ( A \pm \xi A \xi),
\end{align}
where we have the guarantee that $A_+$ commutes with $\xi$ and $A_-$ anti-commutes with $\xi$. Thus, we can read off the the procedure for constructing $M$. We first choose a random matrix $A$ with $A_{jk}\in [-1, 1]$. Next we construct $A_\pm$ using \cref{eqn:Apm}. Finally we construct $M$ according to
\begin{align}
    M = \begin{cases}
        A_+ +iA_- \textup{ for } s=+1, \\
        A_- +iA_+ \textup{ for } s=-1. \\
    \end{cases}
\end{align}

\section{Comparison of spin character at \texorpdfstring{$z = 0$}{z=0} and \texorpdfstring{$z = L$}{z=L}}\label{apx:top_bottom}

\begin{figure*}
    \centering
    \includegraphics[]{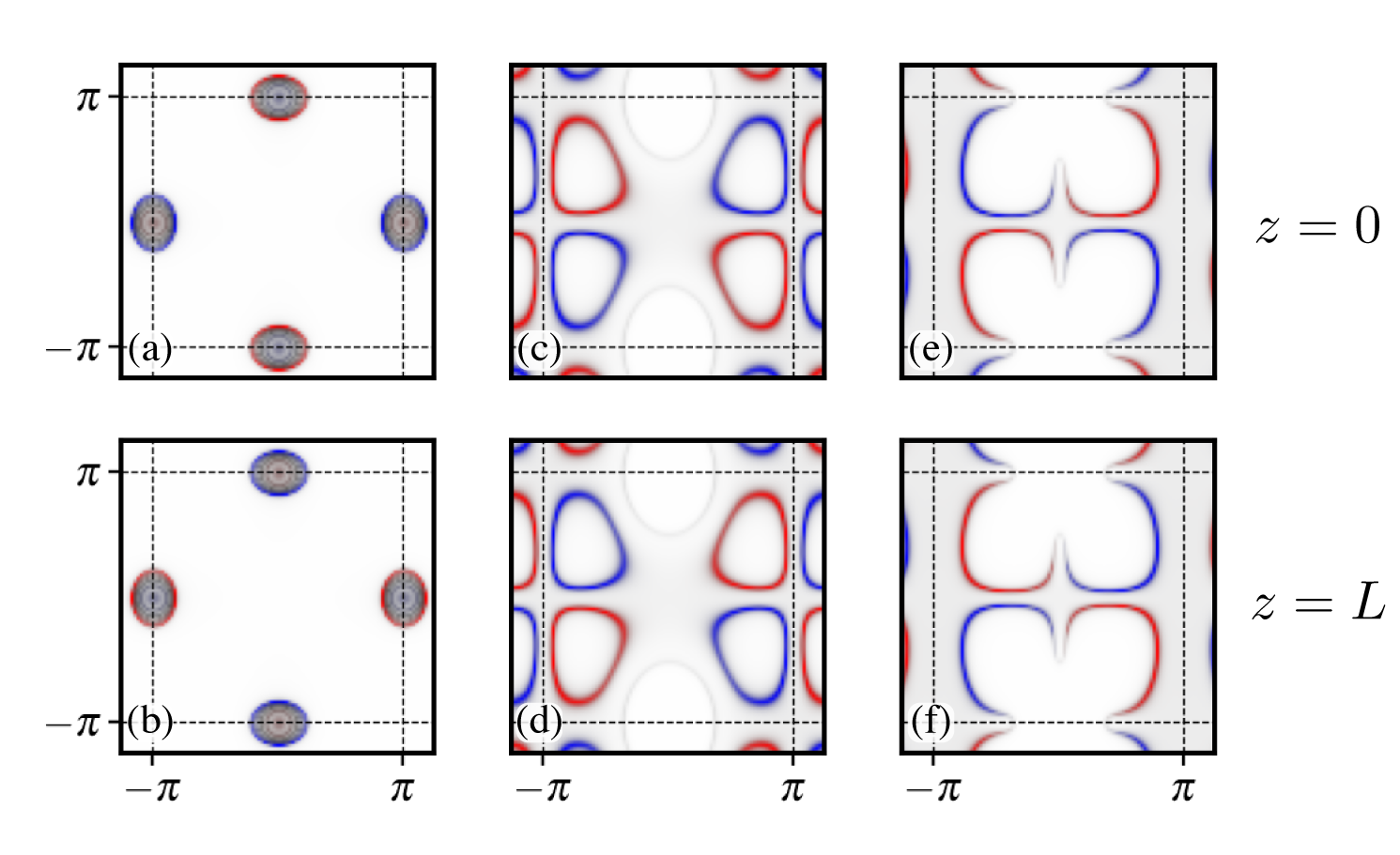}
    \caption{
    The spin spectral density $\mathcal A_s(z)$ for each model is shown for $z = 0$ and $z=L$, corresponding to the top and bottom pairs of layers for each model.
    (a-b) The DLKK-based model.
    (c-d) The nodal-line semimetal.
    (e-f) The Weyl semimetal.}
    
    \label{fig:top_bottom}
\end{figure*}

Here we verify that the spin splitting found at $z=0$ is inverted with respect to the splitting at $Z = L$, which is consistent with the global inversion times time reversal symmetry obeyed by all three models considered, $[\mathcal T \lline \mathcal {TI}]$. In \cref{fig:top_bottom} we show the layer-projected spectral density for the top ($z=0$) and bottom ($z=L$) pairs of layers for the three models considered in this study. In each case we see that the spectral function is identical up to a spin flip.

\end{document}